\documentclass{article}
\usepackage[titletoc]{appendix}
\usepackage{titlesec}
\usepackage[utf8]{inputenc}
\usepackage{tcolorbox}
\usepackage{authblk}
\usepackage{amsmath}
\usepackage{amssymb}
\usepackage{cancel}
\usepackage{diagbox}
\usepackage{caption}
\usepackage{subcaption}
\usepackage{color}
\usepackage{makecell}
\usepackage{hyperref}
\usepackage{tikz}

\newcommand{\barB} {B^-} 
\usepackage{array} 

\usepackage[
backend=biber,
style=numeric,
sorting=none,
maxnames=10, 
maxbibnames=10, 
minnames=10
]{biblatex}
\title{Numerical asymptotics of near-axis expansions of quasisymmetric magnetohydrostatic equilibria with anisotropic pressure}

\author[1,2]{Lanke Fu}
\author[3]{Eduardo Rodriguez}
\author[4]{Rory Conlin}
\author[2]{Amitava Bhattacharjee}
\affil[1]{Princeton Plasma Physics Laboratory, Princeton, NJ 08540, United States}
\affil[2]{Department of Astrophysical Sciences, Princeton University, Princeton, NJ 08544, United States}
\affil[3]{Max Planck Institute for Plasma Physics, Greifswald 17491, Germany}
\affil[4]{Institute for Research in Electronics and Applied Physics, University of Maryland, College Park, MD 20742, United States}
\begin{document}

\maketitle
\section{Abstract} 

Quasisymmetry (QS) is a property of special magnetic configurations, where the magnetic field strength, but not necessarily the full vector field, has a direction of symmetry. QS leads to reduced neoclassical transport and thus can be a desirable property in stellarator design. The Garren-Boozer (GB) conundrum \cite{garren_quasihelical_1991} has been interpreted to mean that globally quasisymmetric magnetohydrostatic (MHS) equilibria, other than axisymmetric solutions, with isotropic pressure do not exist. When expanded as power series of an effective minor radius, the governing equations become overdetermined at the 3rd order. Despite this, recent optimization efforts have found numerical isotropic-pressure equilibria with nearly exact global QS \cite{landreman_optimization_2022, QS_bootstrap, wechsung_precise_2022}. To reconcile these two perspectives, Rodriguez and Bhattacharjee (RB) showed that by introducing pressure anisotropy into the problem, one can overcome the GB conundrum \cite{rodriguez_generalized_boozer_coord, rodriguez_mhd_1}. This formally enables the study of equilibria with exact, global QS. Building on RB's work, we present \texttt{pyAQSC}, the first code for solving the near-axis expansion (NAE) of anisotropic-pressure quasisymmetric equilibria to any order. As a demonstration, we present a 6th order, QA near-axis equilibrium with anisotropic pressure, and a convergence analysis. \texttt{PyAQSC} opens the door to the study of higher-order properties of equilibria with exact global QS. Like existing isotropic-pressure NAE codes, \texttt{PyAQSC} can accelerate stellarator optimization as an initial state tool. However, by optimizing for low pressure anisotropy in a space that allows anisotropy, \texttt{pyAQSC} may discover practical QS stellarator designs previously hard to access. We give results comparing the RB method with DESC \cite{dudt_desc_2020} equilibria with anisotropic pressure.

\section{Introduction}

Quasisymmetry (QS) is a property of magnetic configurations that grants the magnetic field strength, but not necessarily the full vector field, a direction of symmetry \cite{nuhren1988,boozer1983,rodriguez_necessary_2020}. Quasisymmetric stellarators, along with other types of omnigenous stellarators, can achieve neoclassical transport comparable to that of a tokamak. However, unlike tokamaks, these stellarators can generate rotational transform without needing a significant plasma current \cite{helander_theory_2014}. These properties make a quasisymmetric stellarator an attractive magnetic confinement concept. 

Quasisymmetric magnetohydrostatic (MHS) equilibria are notoriously hard to construct. The well-known Garren-Boozer (GB) conundrum has often been interpreted to mean that isotropic-pressure equilibria with global QS may not exist. This is shown using the method of near-axis expansion (NAE) \cite{garren_quasihelical_1991}, which expands the QS and force-balance conditions as power series of an effective minor radius. Generally, the NAE problem becomes overdetermined at the 3rd order \cite{garren_quasihelical_1991, garren_field_strength_1991}, and without special axis shapes, it often fails at the 2nd \cite{landreman_nae}. However, over-determination does not strictly forbid the existence of special, exact solutions. After all, equilibria with exact symmetry are also quasisymmetric. Indeed, recent studies have also found non-axisymmetric numerical equilibria with excellent approximation to QS by optimization \cite{landreman_optimization_2022}. This has led to a focus on obtaining approximate quasisymmetric equilibria by optimization. 

The optimization approach produces quasisymmetric equilibria by iterating a general three-dimensional (3D) equilibrium to minimize its deviation from QS, often alongside other regularization and physics objectives. However, as a nonlinear, non-convex optimization problem, this approach is costly and sensitive to the initial conditions. As a result, quasisymmetric equilibrium optimization often requires long computation time and careful tuning. These challenges have led to a revived interest in NAE as a numerical tool to directly construct configurations with approximate QS \cite{direct_1, direct_2, landreman_nae} or other types of omnigeneity \cite{direct_3,camacho-mata-2022,rodriguez_near-axis_2024,jorge2022c}. A near-axis quasisymmetric equilibrium has fewer parameters than its global equilibrium counterpart. Yet, it can reproduce many important features targeted in optimizations \cite{landreman_nae_optimization_consistency_2019, thesis}. These properties make NAE an effective low-dimensional representation of an approximate quasisymmetric equilibrium. As a result, NAE offers a dramatic speed-up and increased control compared to traditional optimization \cite{landreman_nae, landreman_nae_optimization_consistency_2019, direct_2}. Quasisymmetric optimizations initialized with near-axis solutions often converge faster than those using tokamak-like initial states \cite{giuliani_direct_2023}. This makes NAE the ideal tool for large-scale surveys of QS and omnigenous configuration spaces \cite{Rodríguez_2023_param_space, Rodríguez_2022_phase_space, landreman_mapping_2022}. 

Despite these successes, an isotropic pressure NAE is still fundamentally limited by the GB conundrum. The early truncation limits our knowledge of the truncation error and the geometric complexity of configurations that a NAE can represent \cite{kuo1987, ruth_high-order_2024}. In most cases, it also prohibits the study of important, high-order properties, such as the magnetic shear \cite{rodriguez_magnetic}. These limitations have raised interest in an NAE that continues to arbitrarily high order without over-determination.

In \cite{rodriguez_magnetic, rodriguez_mhd_1, rodriguez_mhd_2}, Rodriguez and Bhattacharjee (RB) showed that the introduction of pressure anisotropy allows a NAE to circumvent the GB conundrum. This enables the construction of equilibria with weak global QS to arbitrarily high order. Although isotropic pressure is a good approximation to the conditions in a fusion plasma, strictly enforcing it as a constraint can make promising QS equilibria inaccessible. However, RB only presented an outline of the expansion procedure, along with a second-order circular-axis example. To systematically study the high-order behavior of the expansion and utilize the expansion as an effective tool for stellarator design, an efficient numerical implementation is necessary.
 
Based on RB, we develop the theory and a computer code to systematically study the high-order behavior of the expansion and apply this to stellarator design. We present \texttt{ piAQSC}, the first code for the NAE of quasisymmetric anisotropic pressure equilibria to any order. \texttt{PyAQSC} has potential in both theoretical and engineering studies. As the first arbitrary-order finite-pressure NAE, it enables the study of MHS equilibria with exact, global QS. As the first anisotropic-pressure NAE code, it allows access to a larger solution space inaccessible to existing NAE codes. As a result, \texttt{pyAQSC} can help discover promising new solutions with good QS and low pressure anisotropy. It may also allow the optimizer to pass through regions of the solution space that were previously prohibited, improving the convergence when optimizing near-axis equilibrium. \texttt{PyAQSC} is fully auto-differentiable, allowing fast, gradient-based optimization of all near-axis quantities. \texttt{PyAQSC} runs on both CPU and GPU, and provides interfaces to \texttt{DESC} and \texttt{Simsopt} \cite{dudt_desc_2020,Landreman2021simsopt}, two established stellarator optimization codes. This makes it simple to integrate into existing optimization scripts on any computer architecture.   

This paper is organized as follows. Section~\ref{section:governing} introduces the governing equations defining our problem of interest as given in \cite{rodriguez_mhd_1}, emphasizing the physical meaning of quantities and equations. Section~\ref{section:expansion} introduces the concept of the NAE, and performs the expansion in effective minor radius $\epsilon$ to yield an ordered equation set. Section~\ref{section:recursion} derives the exact expressions that \texttt{pyAQSC} evaluates to solve the governing system of equations. Section~\ref{section:numerical} summarizes the numerical methods used in \texttt{pyAQSC}. The last part of the paper, Section~\ref{section:validation}, presents an example 6th-order, QA equilibrium generated by \texttt{pyAQSC}, along with validation data for \texttt{pyAQSC}'s accuracy. The estimated radius of convergence corresponds to a small volume around the axis. However, expanding the volume by optimizing the axis shape and profiles may be possible. To demonstrate \texttt{pyAQSC}'s usefulness as a tool for directly constructing QS equilibria, we also present a general 3D equilibrium with good approximate QS by fitting this near-axis solution in \texttt{DESC}\cite{panici_extending_2025}.

\section{Governing equations} 
\label{section:governing}
We start with a summary of the governing equations, obtained by RB, in \cite{rodriguez_mhd_1, rodriguez_mhd_2, rodriguez_magnetic}, which this work implements. The system of governing equations consists of two separate parts. The first part, the magnetic equations, describes a toroidal magnetic field that has nested flux surfaces and is weakly quasisymmetric \cite{rodriguez_necessary_2020}. The second part, the force-balance equation, enforces a MHS equilibrium with parallel and perpendicular pressure. One can choose to solve the full equation set or only the magnetic equations to any order to construct a generic quasisymmetric field or one in anisotropic-pressure MHS equilibrium.

\subsection{The magnetic equations}
\label{section:governing:magnetic}
The magnetic equations describe a weakly quasisymmetric toroidal field with nested flux surfaces \cite{rodriguez_magnetic, rodriguez_mhd_1} :
\begin{gather}
    \begin{split}       
        \textbf{B}&=B_\theta\nabla\chi
            +(B_\alpha(\psi)-\bar{\iota}B_\theta)\nabla\phi
            +B_\psi\nabla\psi\\
        &=\nabla\psi\times\nabla\chi
            +\bar{\iota}\nabla\phi\times\nabla\psi,
    \end{split}\label{eq:original C}\tag{\textbf{C}}\\
    J 
     = \frac{B_\phi + \iota B_\theta}{B^2} = \frac{B_\alpha(\psi)}{B^2},\label{eq:original J}\tag{J}
\end{gather}
where the magnetic coordinates $\{\psi, \theta, \chi\}$ belong to a special class, the Generalized Boozer Coordinate (GBC) \cite{rodriguez_generalized_boozer_coord}.  The co/contravariant equation, \eqref{eq:original C}, describes a solenoidal magnetic field with nested flux surfaces and straight field line coordinates. The Jacobian equation, \eqref{eq:original J}, must be satisfied by a magnetic field with weak QS in GBC. Here, both $B_\alpha$ and $\iota$, the rotational transform, are flux functions. We also define $\chi=\theta-(N/M)\varphi$ and $\bar\iota\equiv\iota-N/M$, where $N/M$ is the helicity of the magnetic axis.  

\subsection{Force balance}\label{section:governing:force balance}
The force balance equation models the balance between the Lorentz force and an anisotropic pressure. Here, the pressure tensor is:
\begin{equation*}
    \begin{split}
        \Pi &= p_\parallel\textbf{bb} 
            + p_\perp (\mathbb{I}-\textbf{bb}) \\
            &= \Delta \textbf{BB} + p_\perp \mathbb{I}.
    \end{split}
\end{equation*}
This pressure tensor makes the distinction between parallel pressure $p_\parallel$, and perpendicular pressure $p_\perp$. Here, $\textbf{b}\equiv \textbf{B}/B$ is the unit vector tangent to  the magnetic field, and $\Delta\equiv (p_\parallel-p_\perp)/B^2$ is the anisotropy.

The MHS force balance equation is then,
\begin{equation}\label{eq:original MHS}
    \textbf{j}\times\textbf{B} = \nabla\cdot\Pi = \nabla\cdot(\Delta\textbf{BB} + p_\perp \mathbb{I}).\tag{\textbf{F}}
\end{equation}
$\{\eqref{eq:original C}, \eqref{eq:original J}, \eqref{eq:original MHS}\}$ constitute the complete set of governing equations. 

\section{Near-axis expansion}
\label{section:expansion}
We now briefly introduce the concept of NAE and summarize the expansion performed by RB. The NAE directly solves for approximate quasisymmetric equilibria by expanding the governing equations as power series in the distance from a known magnetic axis and solving order by order. The expansion parameter used in \cite{rodriguez_magnetic, rodriguez_mhd_1, rodriguez_mhd_2} and this paper is $\epsilon\equiv\sqrt\psi$. To perform this expansion directly, and given that $\psi$ is one of the GBC coordinates \cite{rodriguez_generalized_boozer_coord}, we must rewrite $\{J, \mathbf{C}, \mathbf{F}\}$ in an inverse coordinate form.

\subsection{Inverse coordinate forms}
\label{section:expansion:near-axis form}
To perform expansions in the GBC, we must transform \eqref{eq:original C} and \eqref{eq:original J} into an inverse-coordinate form that uses $\{\psi, \chi, \phi\}$ as the independent variables. Define the inverse transform from GBC to a known orthonormal coordinate, $\textbf{x}(\psi, \chi, \phi)$. Using the dual relations
\begin{equation*}
\frac{\partial\textbf{x}}{\partial\psi}=J\nabla\theta\times\nabla\phi, \quad
\frac{\partial\textbf{x}}{\partial\theta}\times\frac{\partial\textbf{x}}{\partial\phi}=J\nabla\psi \text{, and cyclic permutations,}
\end{equation*}
we can rewrite \eqref{eq:original J} and \eqref{eq:original C} as:
\begin{equation}\label{governing:J vec}\tag{$J$}
    \text{\eqref{eq:original J}} \Leftrightarrow \frac{B^2_\alpha}{B^2} = J^2B^2 = \left| \frac{\partial\textbf{x}}{\partial\phi}+\bar{\iota}\frac{\partial\textbf{x}}{\partial\chi} \right|^2,
\end{equation}
\begin{equation}\label{governing:C vec}\tag{$\textbf{C}$}
    \text{\eqref{eq:original C}} \Leftrightarrow (B_\alpha-\bar{\iota}B_\theta)\frac{\partial\textbf{x}}{\partial\psi}\times\frac{\partial\textbf{x}}{\partial\chi}
    +B_\theta\frac{\partial\textbf{x}}{\partial\phi}\times\frac{\partial\textbf{x}}{\partial\psi}
    +B_\psi\frac{\partial\textbf{x}}{\partial\chi}\times\frac{\partial\textbf{x}}{\partial\phi}
    =\frac{\partial\textbf{x}}{\partial\phi}+\bar{\iota}\frac{\partial\textbf{x}}{\partial\chi}.
\end{equation}
This transforms all terms in the magnetic equations into functions of $\{\psi, \chi, \phi\}$. 

We now introduce a Frenet frame with respect to a known, closed axis $\mathbf{r}_0$, and write $\mathbf{x}$ as:
\begin{equation*}
    \begin{split}
        \textbf{x}(\psi, \chi, \phi) &= \textbf{r}_0[l(\phi)]+X(\psi, \chi, \phi) \hat{\boldsymbol{\kappa}}_0[l(\phi)]+Y(\psi, \chi, \phi) \hat{\boldsymbol{\tau}}_0[l(\phi)]+Z(\psi, \chi, \phi) \hat{\boldsymbol{b}}_0[l(\phi)].
    \end{split}
\end{equation*}
Here, $\hat{\boldsymbol{b}}_0$, $\hat{\boldsymbol{\kappa}}_0$ and $\hat{\boldsymbol{\tau}}_0$ are the tangent, normal, and binormal basis vectors of the Frenet-Serret frame and $X(\psi=0)=Y(\psi=0)=Z(\psi=0)=0$. Here, $l$ is the length along the magnetic axis. Due to the choice of $\phi$ in GBC, $dl/d\phi$ is a constant. The basis vectors satisfy:
\begin{alignat*}{2}
    &\frac{d\textbf{r}_0}{dl}=\hat{\boldsymbol{b}}_0,
    &&\frac{d\hat{\boldsymbol{b}}_0}{dl}=\kappa(l)\hat{\boldsymbol{\kappa}}_0,\\
    &\frac{d\hat{\boldsymbol{\tau}}_0}{dl}=\tau(l)\hat{\boldsymbol{\kappa}}_0,
    &&\frac{d\hat{\boldsymbol{\kappa}}_0}{dl}=-
    \kappa(l)\hat{\boldsymbol{b}}_0-\tau(l)\hat{\boldsymbol{\tau}}_0,
    \\
\end{alignat*}
where $\kappa(l)$ and $\tau(l)$ are the curvature and torsion of the axis. When solving the governing equations, we consider $X$, $Y$ and $Z$ as unknowns, along with other parameters of the magnetic configuration/equilibrium. This allows us to self-consistently construct the magnetic coordinates along with the equilibrium. (Note that the sign of $\tau(l)$ in this paper is the opposite of that in \cite{direct_2}.) In \texttt{pyAQSC}, the axis $\textbf{r}$ will be parameterized in a cylindrical coordinate $\{R, \Phi, Z\}$. In the rest of the paper, the uppercase $\Phi$ will denote the cylindrical toroidal angle, and the lowercase $\phi$ will denote the GBC coordinate.   

Applying this definition of $\textbf{x}$ to \eqref{eq:original C} and \eqref{eq:original J} gives the final form of the magnetic equations used in \texttt{pyAQSC}. To perform the expansion, we project \eqref{eq:original C} onto the Frenet basis, $\{\hat{\boldsymbol{b}}_0, \hat{\boldsymbol{\kappa}}_0, \hat{\boldsymbol{\tau}}_0\}$, and two additional vectors, $\frac{\partial \textbf{x}}{\partial \psi}, \frac{\partial \textbf{x}}{\partial \chi}$. We refer to the resulting scalar equations as $C_b, C_\kappa, C_\tau, C^{\partial\textbf{x}/\partial\psi}$ and $C^{\partial\textbf{x}/\partial\chi}$. Note that $C_b, C_\kappa$ and $C_\tau$ are independent, but $C^{\partial\textbf{x}/\partial\psi}$ and $C^{\partial\textbf{x}/\partial\chi}$ are not. For their full expressions, see Appendix \ref{appendi:govern}.

To expand the force balance equation, we express $\mathbf{j}$ using the covariant form of the magnetic field. We then project \eqref{eq:original MHS} along $\nabla\psi$, $\nabla\chi$, and $\nabla\phi$. We refer to the three resulting scalar equations as $\mathrm{I},\mathrm{II}$ and $\mathrm{III}$. For their full expressions, see Appendix \ref{appendi:govern}.

Thus far, we have converted all governing equations into a system of seven scalar equations that uses $\{\psi, \chi, \phi\}$ as the independent variables: $\{J, C_b, C_\kappa, C_\tau, \mathrm{I},\mathrm{II},\mathrm{III}\}$. The full equation set represents an ideal MHD equilibrium with global weak QS and anisotropic plasma pressure. The magnetic equations, $\{J, C_b, C_\kappa, C_\tau\}$, represent a toroidal field with global weak QS and independent of the force balance. Both sets can be solved by series expansion to any order.

\subsection{Quantities and expansion}
\label{section:expansion:expansion of quantities}
We now expand all functions that depend on $\psi$ as power series of the effective minor radius $\epsilon\equiv\sqrt{\psi}$. Here, for simplicity, we do not normalize $\epsilon$ as in \cite{garren_quasihelical_1991}. Doing so would simply require a trivial rescaling of the results. 

There are seven 3D fields in the governing equations: the coordinate transformation functions, $\{X, Y, Z\}$, the magnetic field covariant components, $\{B_\theta, B_\psi\}$, and the pressure components, $\{p_{\perp}, \Delta\}$. Following \cite{rodriguez_magnetic} and \cite{rodriguez_mhd_1}, and the original \cite{garren_field_strength_1991}, we expand these fields as power-Fourier series with $\phi$-dependent coefficients,
\begin{alignat}{2}\label{expansion:3D}
    X(\psi, \chi, \phi_j) &= \sum_{n=0}^\infty \epsilon^n &&X_n(\chi, \phi_j) \\
    &= \sum_{n=0}^\infty \epsilon^n \sum_{m=-n, even/odd}^n &&X_{n,m}(\phi_j)e^{im\chi},
\end{alignat}
where the Fourier series only contain even or odd mode numbers smaller than $n$ for analyticity near the magnetic axis \cite{kuo1987}. Unlike the customary approach that expresses the $\chi$ dependence in trigonometric Fourier series, here we use complex exponentials. Although this introduces additional computation cost, operations such as convolution and differentials become substantially simpler to implement.

The 2D field $\barB(\psi, \chi)\equiv1/B^2(\psi, \chi)$ is expanded as a power-Fourier series with constant coefficients (due to QS):
\begin{alignat}{2}\label{expansion:2D}
    \barB_{}(\psi, \chi) = &\sum_{n=0}^\infty \epsilon^n &&\barB_{n}(\chi)\\
    &\sum_{n=0}^\infty \epsilon^n \sum_{m=-n, even/odd}^n &&\barB_{n,m}e^{im\chi}
\end{alignat}
Note that in \cite{rodriguez_magnetic, rodriguez_mhd_1, rodriguez_mhd_2}, this expansion coefficient was called $B_{n, m}$. Here, to help avoid confusion with the magnetic field strength $B$, we changed the notation to $\barB_{n,m}$.

The flux functions $B_\alpha$ and $\bar{\iota}$ are expanded as even power series:
\begin{gather}\label{expansion:1D}
    B_\alpha(\psi) = \sum_{n=0}^\infty B_{\alpha,n}\epsilon^{2n},\\
    \bar\iota(\psi) = \sum_{n=0}^\infty \bar\iota_n\epsilon^{2n}
\end{gather}
Substituting \eqref{expansion:3D} - \eqref{expansion:1D} into the governing equations and matching $\epsilon^n$ terms yields an ordered equation set, $\{J_n, \textbf{C}_n, \mathrm{I}_n, \mathrm{II}_n, \mathrm{III}_n\}$. 

\section{Recursion relations}\label{section:recursion}
We now summarize the iteration procedure that \texttt{pyAQSC} follows to solve the resulting hierarchy of equations order-by-order. The procedures outlined in this section only apply to order $n>1$. For the special procedure for the leading orders, see Appendix~\ref{appendix:leading}. This section is intended as a technical reference for future users and developers of \texttt{pyAQSC}. For brevity, we also omit the explicit expressions that \texttt{pyAQSC} evaluates. For these expressions, see Appendices \ref{appendix:recursion:magnetic} and \ref{appendix:recursion:MHD}. 

This section considers two separate problems. Solving the magnetic equations, $\{J_n, \textbf{C}_n\}$, constructs a globally quasisymmetric magnetic field without any assumptions on the nature of the underlying force balance \cite{rodriguez_magnetic}. Solving the full equilibrium problem, $\{J_n, \textbf{C}_n, \mathrm{I}_n, \mathrm{II}_n, \mathrm{III}_n\}$, constructs a globally QS equilibrium with anisotropic pressure \cite{rodriguez_mhd_1, rodriguez_magnetic}.

\subsection{Magnetic recursion relations}\label{section:recursion:magnetic}
We first present the iteration procedure for solving the magnetic equations, $\{J_n, \textbf{C}_n\}$, to obtain $\{X_{n}, Y_{n, m>0}, Z_{n}, B_{\psi n-2, m>0}\}$, as discussed in \cite{rodriguez_magnetic}. Solving these equations iteratively constructs a weakly quasisymmetric magnetic field with no assumptions regarding the underlying force balance.

At each order $n$, the iteration requires the following inputs (or degrees of freedom):
\begin{gather*}
    \{B_{\theta n,m}(\phi) \text{ (includes $\bar{B}_{\theta n,0}$)}, \\
    B_{\psi n-2,0}(\phi), Y_{n,0}(\phi),\bar{Y}^c_{n,1}\\
    \barB_{n,m}, B_{\alpha(n-1)/2\text{ or }(n-2)/2}, \\
    \bar\iota_{(n-1)/2\text{ or }(n-2)/2}\}.
\end{gather*}
In addition to these quantities, the shape of the axis, $R(\Phi)$ and $Z(\Phi)$, must be provided at the start of the construction. The above inputs include all components of the QS magnetic field magnitude, the rotational transform and $B_\alpha$ profiles, as well as the covariant component $B_\theta$, related to toroidal current. For an order-$n$ expansion, the total required inputs consist of $\frac{(n-1)n}{2}+2\lfloor\frac{n}{2}\rfloor+3$
free periodic 1D functions and $\frac{(n+1)(n+2)}{2}+2\lfloor\frac{n+1}{2}\rfloor$ free scalars. 

At each order, we evaluate the unknowns in the following order:
\begin{enumerate}
    \item \underline{$B_{\psi n-2}$}: We can obtain $B_\psi$ by eliminating $Z_n$ from Eqs.~\eqref{governing:Cdxdchi} and \eqref{governing:Cdxdpsi} and obtain a linear ordinary differential equation (ODE) in $\chi$ for $B_{\psi n-2}$, Eq.~\eqref{recursion:B_psi}. This ODE can be solved exactly, at odd orders giving the full value of $B_{\psi n-2}$, but leaving $B_{\psi n-2,0}$ out of the $m=0$ component at even orders. Because of this, $B_{\psi n-2, 0}$ is free, and the equation needs to be used to constrain $Y_{n, 1}$ instead (see step 4). 

    \item \underline{$Z_n$ and $B_{\theta n,\pm n}$}: $Z_n$ can be algebraically calculated from Eq.~\eqref{recursion:Z}, derived by summing $C_{\tau}^{n-1}$ and $ C_{\kappa}^{n-1}$. In addition to $Z_n$, Eq.~\eqref{recursion:Z} also forces $B_{\theta n,\pm n}$ to zero. 
    The solution for $Z_n$ involves $B_{\psi n-2}$ from step 1.
    
    \item \underline{$X_n$}: we calculate $X_n$ algebraically from Eq.~\eqref{recursion:X}, derived from the $n$-th order Jacobian equation, $J_n$. This involves directly $Z_n$ from step 2.

    \item \underline{$Y_n$}: we obtain all but one $\chi$ Fourier components of $Y_n$ from a linear first-order ODE in $\chi$, Eq.~\eqref{recursion:Y recursion}, derived from $C_b^{n-1}$. This crucially involves $X_n$ from step 3 directly. 
    The ODE has a unique periodic solution for all $B_{\alpha0}, X_1\neq0$, and has a free Fourier coefficient in $\chi$, $Y^\text{free}_{n}(\phi)$. At even orders, we choose $Y_{n, 0}$ as the free component, which we shall treat as an input. At odd orders, we choose $Y_{n, 1}$ as the free component, which must however satisfy the linear ODE Eq.~\eqref{recursion:Y ODE} mentioned in step 1. 
\end{enumerate}

For the exact formulae of the recursion relations, see Appendix~\ref{appendix:recursion:magnetic}.

\subsection{Full-equilibrium recursion relations}\label{section:recursion:MHD}
The force balance conditions substantially modify the iteration procedure. To construct QS, anisotropic pressure equilibria, we must solve for
$$
\{B_{\theta n}, B_{\psi n-2}, p_{\perp n}, \Delta_n, X_n, Y_n, Z_n\},
$$
using the following inputs (degrees of freedom): 
\begin{gather*}
    \{\bar{B}_{\theta n,0}, \barB_{n}, p_{\perp0,0}(\phi), \bar\Delta_0, \\
    B_{\alpha(n-1)/2\text{ or }(n-2)/2},\\
    \bar\iota_{(n-1)/2\text{ or }(n-2)/2} \text{ or } \bar B_{\theta 2k, 0},\\
    R(\Phi), Z(\Phi)\}.
\end{gather*}
The barred quantities here denote averages over the toroidal angle $\phi$.

When solving the full equation set, $\{B_{\theta n}, B_{\theta n+1,0}(\text{odd }n), B_{\psi n,0}(\text{even }n), Y^\text{free}_{n}\}$ must be solved together self-consistently in a linear, inhomogeneous ODE system, called the looped equations. The iteration procedure is highly coupled and must be treated two orders at a time. Order $2k-1$ and $2k$ combined, the required inputs include 6 free periodic 1D functions and $4k^2 + 8k + 5$ free scalars. Appendix \ref{appendix:looped} details the derivation of the looped equations. 


For consecutive odd and even orders, $2k-1$ and $2k$, the procedure can be summarized as follows:
\begin{enumerate}

\item  \underline{$B_{\theta 2k-1}$, $B_{\theta 2k,0}$ and $Y_{2k-1,1}$}: We start by solving the odd-order looped equations, Eqs.~\eqref{governing:MHD II tilde} and \eqref{recursion:Y ODE}. Both together constitute a system of $2k$ linear inhomogeneous ODEs on the unknowns $\{B_{\theta 2k-1}, B_{\theta 2k,0}, Y_{2k-1,1}\}$ in $\phi$. The required inputs are all order $2(k-1)$ unknowns, $B^-_{2k-1}$, and crucially one of the following two scalar constants: $\bar\iota_{k-1}$ or $\bar B_{\theta 2k, 0}\equiv\oint d\phi B_{\theta 2k,0}$.\footnote{RB showed that it is also possible to solve this system relaxing one of $\bar{\iota}_{(n-1)/2}$ or $\bar B_{\theta n+1,0}$, and providing a third scalar $ \sigma_n(0)\equiv Y^c_{n,1}(\phi=0)/Y^s_{n,1}(\phi=0)$, related to stellarator symmetry breaking. This is not currently supported by \texttt{pyAQSC}\cite{rodriguez_mhd_1}.}
 
\item \underline{$B_{\psi 2k-3}$, $Z_{2k-1}$, $X_{2k-1}$ and $Y_{2k-1}$}: we obtain $B_{\psi 2k-3}$, $Z_{2k-1}$, $X_{2k-1}$ and $Y_{2k-1}$ following steps 1 through 4 of the previous section, except that the ODE for $Y_{2k-1,1}$ has become a part of the looped equations (and actually $B_{\theta 2k-1}$ is no longer an input).

\item \underline{$p_{\perp 2k-1}$ and $\Delta_{2k-1}$}: $p_{\perp 2k-1}$ is given algebraically by Eq.~\eqref{recursion:p}, derived from Eq.~\eqref{governing:MHD III}. The anisotropy $\Delta_{2k-1}$ is given by the linear inhomogeneous PDE Eq.~\eqref{recursion:Delta}, derived from \eqref{governing:MHD I}. 

\item  \underline{$B_{\theta 2k}$, $B_{\psi 2k-2,0}$, $Y_{2k,0}$ and $\bar\Delta_{2k,0}$}: we solve the even-order looped equation for $\{B_{\theta 2k}, B_{\psi 2k-2,0}, Y_{2k,0}\}$, using quantities from order $2k-1$, including $B_{\theta 2k,0}$. The solution also constrains $\bar\Delta_{2k,0}\equiv\oint d\phi \Delta_{ 2k,0}$, the average anisotropy, to be consistent with a periodic $B_{\psi 2k-2,0}$. 

\item \underline{$B_{\psi 2k-2}$, $Z_{2k}$, $X_{2k}$ and $Y_{2k}$}: once again, we repeat steps 1 through 4 in the previous section to obtain $B_{\psi 2k-2}$, $Z_{2k}$, $X_{2k}$ and $Y_{2k}$. As in step 2, the ODE for $Y_{2k,0}$ becomes part of the looped equations (and $B_{\theta 2k}$ and $B_{\psi 2k-2, 0}$ are no longer inputs).

\item \underline{$p_{\perp 2k}$ and $\Delta_{2k}$}: this step is analogous to step 3. $p_{\perp 2k}$ is given algebraically by Eq.~\eqref{recursion:p}, derived from Eq.~\eqref{governing:MHD III}. $\Delta_{2k}$ is given by the linear inhomogeneous PDE Eq.~\eqref{recursion:Delta}, derived from \eqref{governing:MHD I}. The even-order PDE for $\Delta_{2k}$ has a unique solution up to $\bar\Delta_{2k,0}$, which was found in step 4.
\end{enumerate}

For the explicit recursion relations and equation counting, see \ref{appendix:recursion:MHD} and \ref{appendix:looped}.

\section{Implementation}\label{section:numerical}
\texttt{PyAQSC} constructs quasisymmetric magnetic fields by evaluating the recursion relations summarized in Section~\ref{section:recursion}. In this section, we present the key numerical methods and packages used in the implementation of \texttt{pyAQSC}, as well as its notable features.

\texttt{pyAQSC} stores the value of $\phi$-dependent quantities on a uniformly spaced grid, and uses pseudo-spectral methods for the numerical evaluation of integrals, derivatives, and linear ODEs. See Appendix \ref{appendix:spectral} for the detailed implementation. The leading orders contain a periodic Riccati equation with an unknown scalar constant \cite{direct_2}. \texttt{PyAQSC} solves this equation using a 4th-order Runga-Kutta (hereafter, RK4) shooting method, as discussed in Appendix~\ref{appendix:leading}. As the expansion is taken to higher orders, repeated $\phi$ derivatives in the recursion relations amplify the high-frequency noise in $\phi$. To alleviate this issue, \texttt{pyAQSC} truncates the $\phi$-Fourier coefficients of each unknown after each order's iteration. The truncation mode number, $M_{\text{max}, n}$, is manually identified by scanning a list of candidate frequencies and choosing a small number with low iteration errors. Future improvements will develop a method to adaptively select the truncation point, or a regularization method that limits high-frequency fluctuations by constraining the inputs \cite{ruth_high-order_2024}.

\texttt{PyAQSC} performs all numerical operations with \texttt{JAX} \cite{jax2018github}, a Python numerical library with both CPU and GPU backends. It can run on a range of HPC platforms with minimal machine-specific setup. \texttt{PyAQSC} contains no symbolic components. Therefore, it can efficiently calculate high-order terms for divergence rate studies, or alternatively, compute a large number of low-order solutions for optimization and configuration space surveys. \texttt{PyAQSC} supports automatic differentiation, which enables fast, gradient-based optimization of all near-axis quantities. \texttt{PyAQSC} provides interfaces with \texttt{DESC} \cite{dudt_desc_2020} and \texttt{Simsopt} \cite{Landreman2021simsopt}, two established stellarator optimization suites. This simplifies the integration of \texttt{pyAQSC}-based initial states and figures of merit into existing optimization codes.

\section{Numerical results}\label{section:validation}
As discussed earlier, NAE expansions for MHS equilibria with isotropic pressure suffer from the overdetermination problem at third and higher orders. In this section, we present the first 6th-order, anisotropic-pressure near-axis equilibrium which demonstrates explicitly the new capabilities made accessible by the RB formulation. We consider a 2-field-period, QA equilibrium with stellarator symmetry. With this equilibrium, we measure the solution's radius of convergence to order 6 and validate that \texttt{pyAQSC} can correctly evaluate the recursion relations. We also compare the near-axis solution with a series of global anisotropic-pressure equilibria that use near-axis flux surfaces with varying aspect ratios as their boundary conditions. Last, we demonstrate \texttt{pyAQSC}'s effectiveness as a tool for directly constructing QS stellarator equilibria. We fit a global equilibrium with the same near-axis behavior as our solution \cite{panici_extending_2025}. Without optimizing for QS, the global fit achieves QS quality comparable to many known optimized equilibria.

\subsection{Parameters}
The axis  in cylindrical coordinate $(R, \Phi, Z)$ is:
\begin{equation}
    \begin{split}
        R &= 1+0.1\cos(2\Phi)\,\text{(m)},\\
        Z &= 0.1\sin(2\Phi)\,\text{(m)}.\\
    \end{split}
\end{equation}
Its on-axis pressure is:
\begin{equation}
    p_0 = \frac{1}{20}[1+0.1\cos(2\phi)]\, (Pa\cdot\mu_0)
\end{equation}
Its components of magnetic field strength are:
\begin{equation}
\begin{split}
       \barB_0& = 1 \left( \text{T}^{-2} \right), \\
       B^-_{1,1c}&=-1.8 \left(  \text{T}^{-5/2}/ \text{m} \right),\\
       B^-_n&=0 \text{ for } n\geq2 
\end{split}
\end{equation}
To reduce anisotropy and toroidal current, we set:
\begin{equation}
    \bar{\Delta}_0, \bar B_{\theta n,0}=0\text{ for } n\geq2.
\end{equation}
For this solution, we use 200 grid points for the $\phi$ dependence. During axis calculations and RK4 solves, we increase the resolution to 5000 points to improve accuracy. Table \ref{tab:low_pass} shows the empirical truncation mode number $M_{\text{max}, n}$ for the low-pass filter. See Appendix \ref{appendix:low_pass} for the selection method for $M_{\text{max}, n}$. 
\begin{table}
    \centering
    \begin{tabular}{|c|c|c|c|c|c|c|} \hline 
         Order&  $n=1$&  $n=2$&  $n=3$&  $n=4$&  $n=5$& $n=6$\\ \hline 
         $M_{\text{max}, n}$&  $45$&  $50$&  $45$&  $40$&  $35$& $30$\\ \hline
    \end{tabular}
    \caption{Empirical truncation mode number used in the low-pass filter. }
    \label{tab:low_pass}
\end{table}

\subsection{Performance}
Table \ref{tab:time} shows the typical run time for each order in \texttt{pyAQSC}, on CPU and GPU. As a reference, existing scalar-pressure NAE codes can run in milliseconds on a single core \cite{landreman_nae}. For a single case, \texttt{pyAQSC} runs in seconds. The CPU run time is slightly shorter than GPU. This is $10^3\times$ slower than existing scalar pressure codes, but still $10^2\sim10^3\times$ faster than global equilibrium solves. 

When performing parameter scans, \texttt{pyAQSC}'s GPU performance gains a $500\times$ speed-up from vectorization using \texttt{jax.vmap}. \texttt{vmap} allows \texttt{pyAQSC} to perform batches of NAE in parallel without repeating certain overheads. When vectorized on GPU, \texttt{pyAQSC} can achieve millisecond-level runtimes on par with scalar pressure NAE codes \cite{landreman_nae}. The CPU performance benefits less from \texttt{vmap}. Vectorization comes at a memory cost. In the present version, $n=5, 6$ on GPU and $n=1,2$ on CPU both cause out-of-memory errors. Nevertheless, we still believe that the GPU vectorization performance at $n \leq 4$ demonstrates the validity of our numerical methods. We hypothesize that the slower single-case runtimes are likely due to overheads, such as internal logic and CPU-GPU communication. 

This paper only presents a first version of \texttt{pyAQSC}. In future developments, we will simplify \texttt{pyAQSC}'s internal logic to improve single-case performance. We will also optimize its memory usage to allow vectorization at higher orders.
\begin{table}
    \centering
    \begin{tabular}{|c|c|c|c|} \hline 
         Order&  $n=1, 2$&  $n=3, 4$& $n=5, 6$\\ \hline 
         \makecell{CPU Run time, \\single case}& $0.89$s.& $0.47$s.&$0.03$s.\\\hline
         \makecell{GPU Run time, \\single case}& $1.95$s.& $2.33$s.&$2.72$s.\\\hline
         \makecell{CPU Run time \\(\texttt{vmap} over 1000 cases)}& \makecell{Fails, requires\\$>300$GB RAM.}& \makecell{$0.31$s, \\$312.49$s in total.}&\makecell{$0.48$s/case,\\$480.81$s in total.}\\\hline
         \makecell{GPU Run time \\(\texttt{vmap} over 1000 cases)}& \makecell{$4.06$ms/case,\\$4.06$s in total.}& \makecell{$1.02$ms/case,\\$1.02$s in total.}&\makecell{Fails, requires\\$>40$GB VRAM.}\\\hline
 \makecell{CPU \texttt{DESC}\\solve time (Fig.\ref{fig:scaling})}& \makecell{$64.73\sim 3026.77$s,\\depends on the \\aspect ratio $A$.}&\makecell{$103.68\sim 3159.52$s,\\depends on $A$}.&\makecell{$72.10\sim3198.70$s,\\depends on $A$.}\\\hline
    \end{tabular}
    \caption{The run time of \texttt{pyAQSC} on 2x AMD EPYC 7763 CPUs, and a Nvidia A100 GPU with $40$GB VRAM. Note the substantial GPU performance gain from vectorization.}
    \label{tab:time}
\end{table}

\subsection{High-order behavior}\label{section:numerical:high-order}
\begin{figure}
    \centering
    \includegraphics[width=1\linewidth]{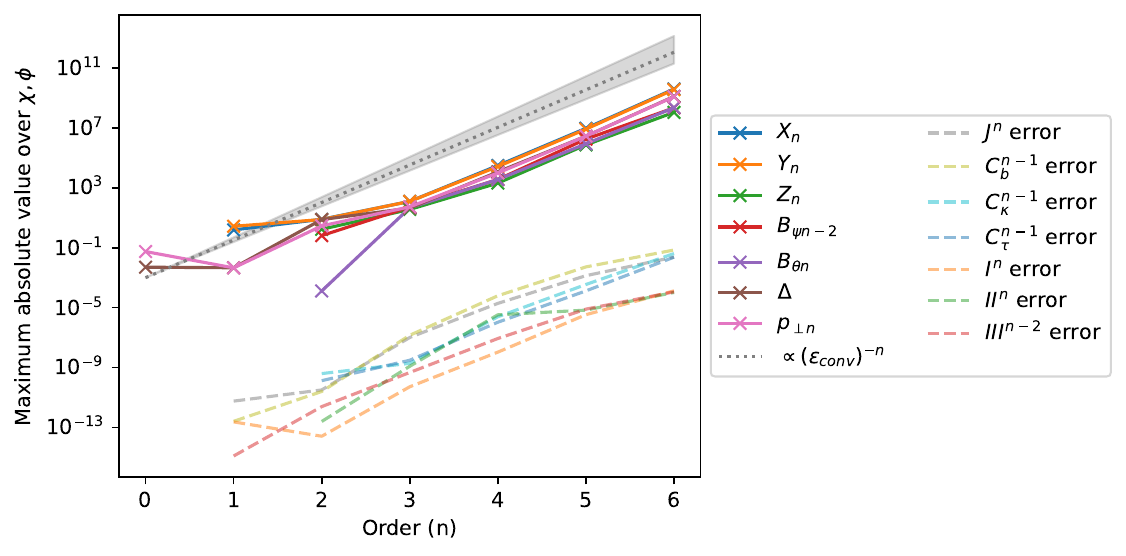}
    \caption{Exponential divergence of the NAE at higher orders. The y axis measures the maximum absolute value of each variable on a $500\times500
$ $(\chi, \phi)$ grid. The dotted line and shadowed area shows the measured radius of convergence, $\psi_\text{conv}=(9.76\pm6.65)\times10^{-6}$Tm$^2$, along with the error interval. The dashed lines shows the maximum absolute values of the errors in each governing equation.}
    \label{fig:exp}
\end{figure}

Fig.\ref{fig:exp} shows the high-order behavior of the expansion. The dashed lines show the measured errors in the ordered governing equations. Comparing the dashed and solid lines, the iteration error is sufficiently low at all orders, confirming that our symbolic order-matching and filtering schemes are accurate. Note that this error measures the inaccuracies during the evaluation of the recursion relations, and is not to be confused with the difference between the NAE and a global equilibrium solver, which we will discuss in depth shortly. At $n\leq3$, most coefficients grow at an increasing rate, and at $n\geq4$, the coefficients appear to grow exponentially at a nearly constant rate. Assuming that all power coefficients at $n\geq4$ grow at an equal rate $\alpha$, and the aspect ratio $A_n\propto\alpha^n$, we measure an average growth rate of $\bar\alpha=\overline{(A_{n+1}/A_{n})}\approx320\pm110$ for $n\geq4$. This gives an approximate radius of convergence $\psi_\text{conv}=1/\bar\alpha^2 \approx(9.76\pm6.65)\times10^{-6}$Tm$^2$. Within the volume bounded by $\psi_\text{conv}$, the 6th order NAE is convergent. The aspect ratio of this flux surface is measured $A_\text{conv}=224$ using DESC. The $A_\text{conv}$ of this configuration is impractically large. However, by optimizing the inputs, it may be possible to expand the volume of convergence and generate globally QS near-axis equilibria with a practical aspect ratio.

\begin{figure}
    \centering
    \subfloat[]{
        \includegraphics[width=0.43\linewidth]{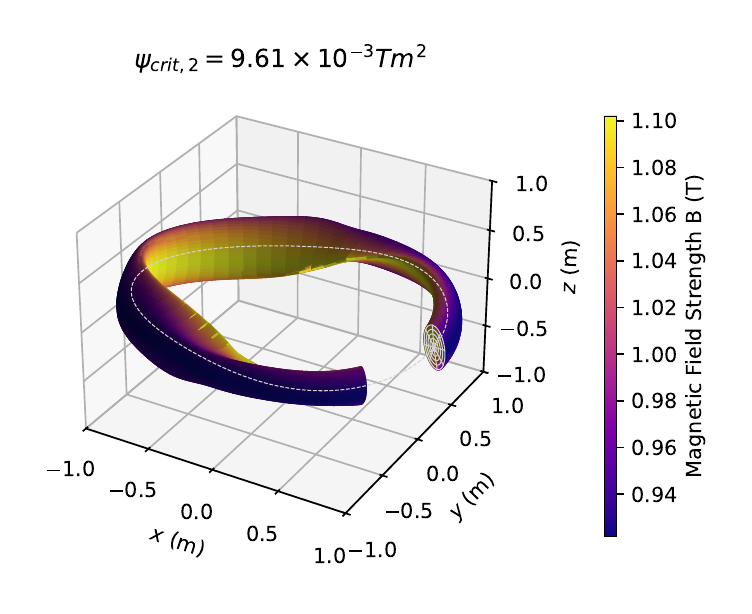}
    }
    \qquad
    \subfloat[]{ 
        \includegraphics[width=0.43\linewidth]{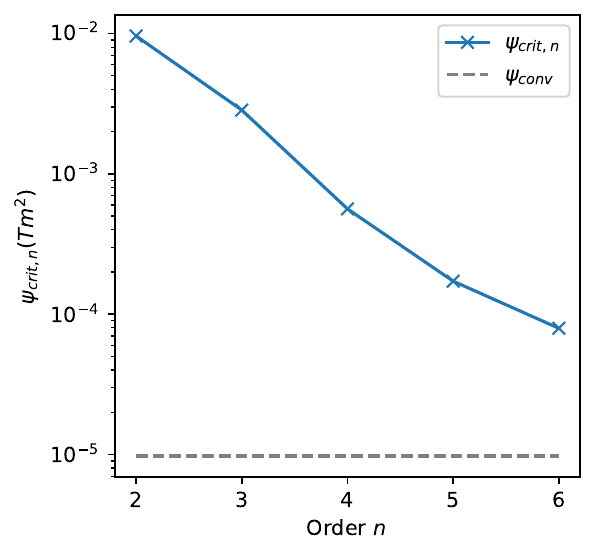}
    }
    \caption{The left shows the last well-behaved flux surface at $n=2$. The right shows the change in $\psi_{\text{crit}, n}$ with increasing $n$. Note the slow convergence of $\psi_{\text{crit}, n}$ to $\psi_\text{conv}$.}
    \label{fig:conv}
\end{figure}

The magnitude of $\psi_\text{conv}$ may be related to the breakdown of the flux surfaces. Similar to scalar-pressure NAEs  \cite{landreman_figures_of_merit_2021}, in \texttt{pyAQSC}, flux surfaces begin to self-intersect when $\psi$ exceeds a threshold, $\psi_\text{crit}$. The threshold $\psi_{\text{crit}, n}$ decreases with $n$. Fig. \ref{fig:conv} shows the values of $\psi_{\text{crit}, n}$ at increasing $n$. Notably, as $n$ increases, $\psi_{\text{crit}, n}$ appears to approach $\psi_\text{conv}$. We hypothesize that $\lim_{n\rightarrow\infty}\psi_{\text{crit}, n}=\psi_\text{conv}$. Intuitively, self-intersections in flux surfaces will break the underlying assumptions of the magnetic equations. This can result in the breakdown of the NAE and the divergence of the power series. The breakdown of the NAE may also indicate island/chaos formation. It is also possible that the divergence is an artifact of the perturbative techniques used, and disconnected with physics. These issues require further investigation. 
\par
To study the difference between a high-order NAE and a global equilibrium solver, we will compare truncated \texttt{pyAQSC} solutions with global anisotropic equilibria from \texttt{DESC}. For this comparison, we obtain 6 near-axis equilibria by truncating the NAE at order $n=1\sim6$. We then generate a set of flux surfaces with increasing $\epsilon$, and decreasing aspect ratio $A$, up to $\psi=\psi_{\text{crit}, n}$. For each flux surface, we solve for the anisotropic equilibrium bounded by this surface, with the same $\{p_\perp(\psi,\chi,\psi), \Delta(\psi,\chi,\psi), \bar\iota(\psi)\}$ profiles as the near-axis equilibrium. Thanks to the similarities in code bases, this can be accomplished using existing near-axis routines in \texttt{DESC}. We define $B_\text{diff}$ as the maximum deviation between the near-axis values of the on-axis magnetic field, $B_\text{axis}^\text{AQSC}$, and its \texttt{DESC} counterpart, $B_\text{axis}^\text{\texttt{DESC}}$. We use $B_\text{diff}$ to measure the error of the near-axis expansion when compared to a global, anisotropic equilibrium. 

Before discussing the error's high-order behavior, we must first clarify the limitations of our method for fitting global equilibria to the \texttt{pyAQSC} solution. \cite{landreman_nae} showed that our method of directly using a near-axis flux surface as the boundary condition in a global equilibrium solver will likely introduce an $O(A^{-n})$ fitting error. The Appendix B of \cite{landreman_nae} presents a remedy for this error in an order-2 isotropic-pressure NAE. However, this treatment is non-trivial to generalize to higher orders. Therefore, we will take this error term into account when discussing the scaling behavior of $B_\text{diff}$ in the following paragraph.

\begin{figure}
    \centering
    \includegraphics[width=1\linewidth]{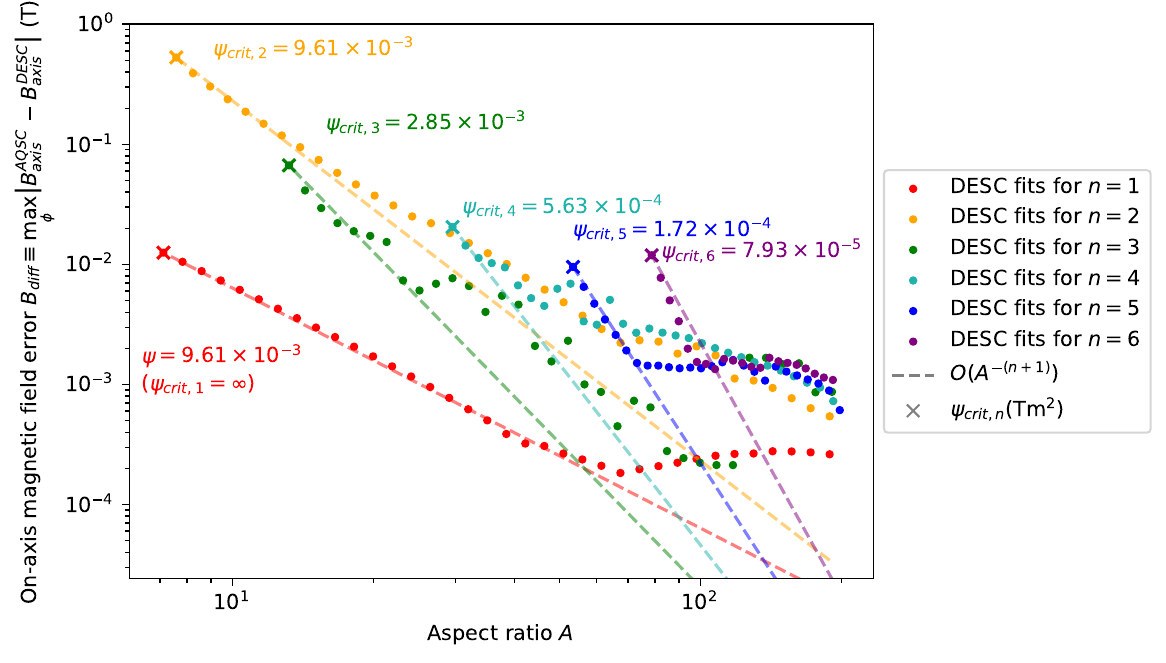}
    \caption{The scaling behavior of \texttt{pyAQSC}. The dots represents \texttt{DESC} equilibria solved using \texttt{pyAQSC} boundary and profiles with varying $n$ and $A$. The dashed lines traces $y=C_nA^{-(n+1)}$. The value of the constant coefficient $C_n$ in each line is chosen so that the line passes through the leftmost dot on each scatter plot, where $\psi_\text{boundary}=\psi_{\text{crit}, n}$.}
    \label{fig:scaling}
\end{figure}

Fig. \ref{fig:scaling} plots $B_\text{diff}$ against $A$. At low $A$ and large $\psi$, we expect $B_\text{diff}$ to scale by $O[A^{-(n+1)}]$, because na\"{i}vely, the error represents the residual of an asymptotic expansion where the higher order modes are large. This trend is clearly visible for $n=1, 2, 3, 5$ and $6$. As $A$ increases, the $O(A^{-n})$ flux surface fitting error of \cite{landreman_nae} will begin to dominate. This trend is clearly visible for $n=2$ and $3$. The scaling behavior for $n=4$ does not appear to agree with expectations. However, because order $5$ and $6$ use information from order $4$ and behave as expected, we believe this deviation is due to numerical errors in the global equilibrium solver. At high $A$, the scaling behavior of $B_\text{diff}$ breaks down across all orders. The onset of this "tail" appears to depend on the resolution choice in \texttt{DESC}, and is likely a result of numerical errors. In conclusion, we believe the scaling behavior of $B_\text{diff}$ has sufficiently validated the accuracy of \texttt{pyAQSC} as an anisotropic-pressure equilibrium solver.

\subsection{Initial states using \texttt{pyAQSC}}
\begin{figure}
    \centering
    \includegraphics[width=0.8\linewidth]{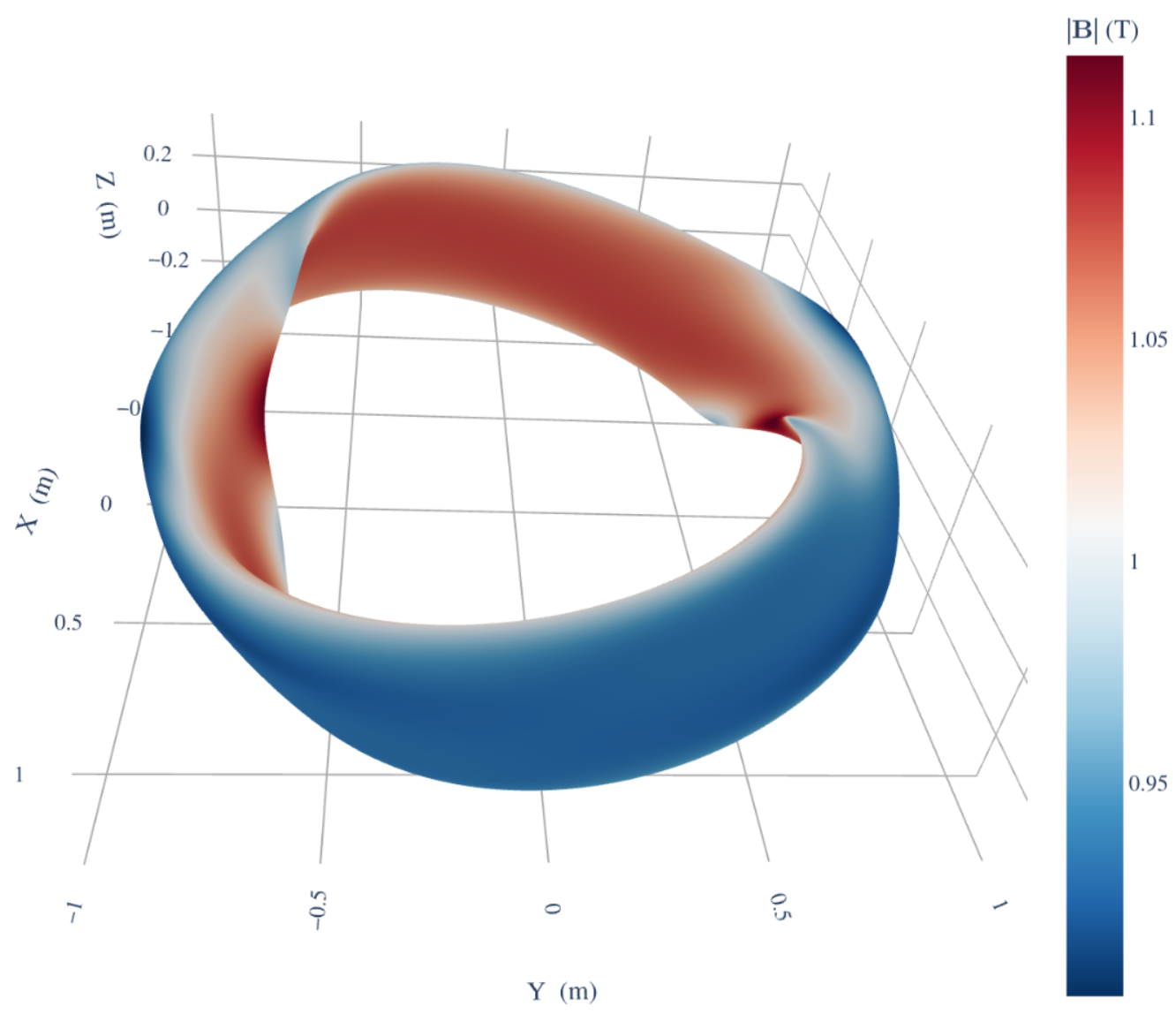}
    \caption{The boundary of the \texttt{DESC} equilibrium, and the magnetic field strength on the boundary.}
    \label{fig:equil}
\end{figure}

To demonstrate the effectiveness of \texttt{pyAQSC} as a tool for directly constructing QS equilibria, we generate a \texttt{DESC} equilibrium with the same near-axis behavior as the \texttt{pyAQSC} solution \cite{panici_extending_2025}. Because $\psi_\text{conv}$ in the above-mentioned equilibrium is impractically small, we choose instead to truncate our expansion at $n=2$, as is commonly practiced with scalar-pressure NAE. This gives a configuration with a moderate aspect ratio of $A=7.5$. Unlike in the previous section, where the DESC equilibrium uses a fixed boundary generated by the NAE, we now constrain the $p_\perp$, $\Delta$, $\iota$ profiles and the asymptotic behavior of the \texttt{DESC} Fourier-Zerneke basis to the near-axis solution, and then solve for minimum force-balance error \cite{panici_extending_2025}. Unlike in Section \ref{section:numerical:high-order}, the boundary shape is now allowed to change to minimize the force balance error. We truncate the equilibrium at $\psi= 0.9\psi_\text{crit}$ to avoid flux surfaces with sharp features or self-intersections. Figure \ref{fig:equil} shows the outer boundary of the resulting \texttt{DESC} equilibrium, and Tab. \ref{tab:properties} lists its global properties. The normalization factor for anisotropic force-balance error in Tab. \ref{tab:properties}, $F_\text{norm}$, is defined based on the magnetic field and volume of the equilibrium:
\begin{equation}
   F_\text{norm}\equiv(1.25)^2\frac{R_{0}}{a^3}\frac{\psi^2}{\mu_0},
\end{equation}
where $R_{0}$ and $a$ are the effective major and minor radii. Note that the procedure in \cite{panici_extending_2025} is designed for isotropic-pressure NAE. The purpose of this section is to demonstrate the effectiveness of \texttt{pyAQSC} as an initial state tool, using existing features in \texttt{DESC}. The development of a fitting procedure for anisotropic pressure is beyond the scope of this paper.
\begin{table}
    \centering
    \begin{tabular}{|l|r|} \hline 
 Average major radius $R_0$&$1.00$ m\\ \hline 
 Average minor radius $a$&$0.134$ m\\ \hline 
         Aspect ratio $A$& $7.49$\\ \hline 
         Volume-averaged $\beta$, $\langle\beta\rangle_\text{vol}$& $10.0\%$\\ \hline 
 Volume-averaged anisotropy, $\left\langle\left|\frac{ \mu_0 (p_{||} - p_{\perp})}{B^2}\right|\right\rangle_\text{vol}$&$1.90\%$\\ \hline 
         Toroidal current $I$& $7802$ A\\ \hline
 Normalized average anisotropic force-balance error $F_\text{error}/F_\text{norm}$.&$6.058\times 10^{-6}$\\\hline
    \end{tabular}
    \caption{Global properties of the \texttt{DESC} fit.}
    \label{tab:properties}
\end{table}

\begin{figure}
    \centering
    \includegraphics[width=0.6\linewidth]{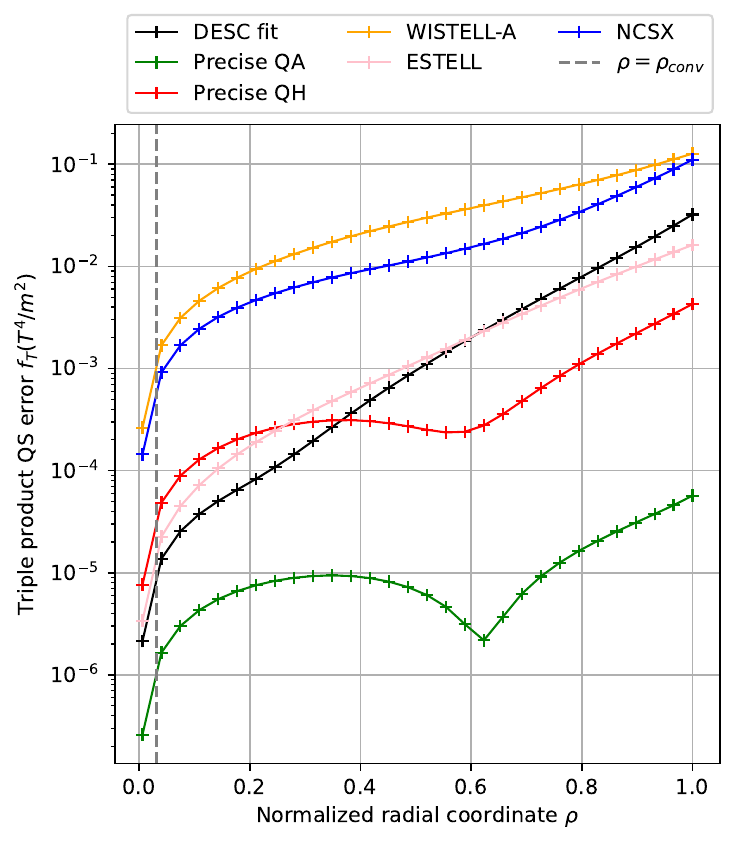}
    \caption{A comparison between the triple product QS error $f_T$ in our \texttt{DESC} fit with other known QS equilibria. The boundary of the convergent volume is marked by the gray dashed line for reference.}
    \label{fig:qs_error}
\end{figure}

\begin{figure}
    \centering
    \includegraphics[width=1.0\linewidth]{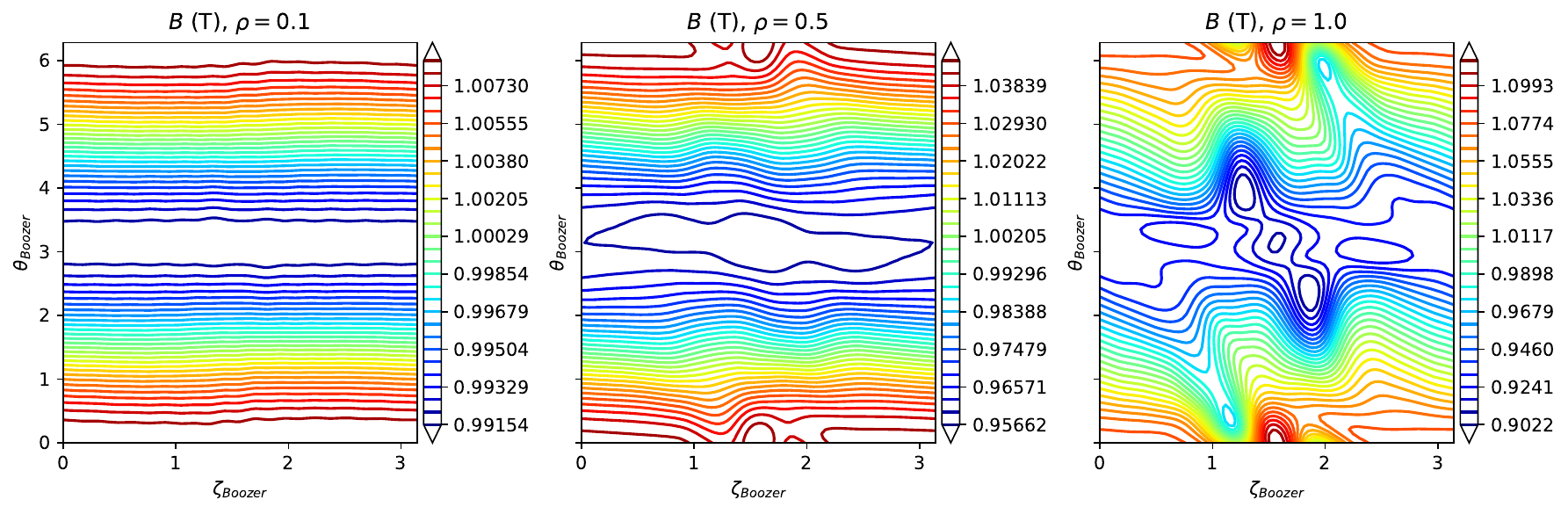}
    \caption{The magnetic field strength contour in the \texttt{DESC} fit. Note the gradual decrease in QS quality with increasing $\rho=\sqrt{\psi/\psi_\text{LCFS}}$.}
    \label{fig:qs_boozer}
\end{figure}

To illustrate a correspondence between the \texttt{pyAQSC} solution space and the space of anisotropic-pressure equilibria with QS, we compare the triple product QS error \cite{helander_theory_2014, rodriguez2022measures} 
\begin{equation}
    f_T\equiv\nabla \psi \times \nabla B \cdot \nabla (\mathbf{B} \cdot \nabla B)
\end{equation}of the \texttt{DESC} "fit" to 6 existing QS configurations: NCSX \cite{nelson_design_2003}, WISTELL-A \cite{bader_advancing_2020}, ESTELL \cite{drevlak_estell:_2013}, Landreman-Paul QH and QA \cite{landreman_optimization_2022}. As Fig.\ref{fig:qs_error} shows, the \texttt{DESC} "fit" has consistently lower QS error than WISTELL-A, NCSX and ESTELL in a large volume, and approaches precise QS near its axis.  Fig. \ref{fig:qs_boozer} shows the magnetic field strength contours in the Boozer coordinate $(\theta_\text{boozer}, \zeta_\text{boozer})$. 

If the $A_\text{conv}$ can be sufficiently increased in future studies, it may be possible to generate a DESC equilibrium with good global QS directly using the convergent volume. This will potentially lead to anisotropic-pressure equilibria with good global force balance and QS.

\section{Conclusions}
In this paper, we presented \texttt{pyAQSC}, the first numerical near-axis expansion code for anisotropic QS equilibria to any order, and the first 6th order near-axis, QA, anisotropic equilibrium. We confirm that \texttt{pyAQSC} can solve the ordered equation sets correctly with low numerical error. Comparing \texttt{pyAQSC} with the global equilibrium solver \texttt{DESC}, the residuals at order $1, 2, 3, 5, 6$ exhibit the expected scaling behaviors. The expansion is only convergent within a small volume around the axis, but this can potentially be improved by optimization. In the future, we will perform parameter space studies similar to \cite{Rodríguez_2022_phase_space, landreman_mapping_2022}, and compare the parameter spaces of \texttt{pyAQSC} with that of isotropic NAE. We will also attempt to search for practical solutions with large $\psi_\text{crit}$, low anisotropy, and practical $A_\text{conv}$, that may serve as the starting points of detailed equilibrium optimization.

\section{Data availability statement}
The data that support the findings of this study are openly available \cite{lanke_pyaqsc_data}.

\titleformat{\section}{\normalfont\large\bfseries}{Appendix \thesection}{1em}{}
\begin{appendices}
\section{Algebraic details}\label{appendix:algebra}

\subsection{Notations}
As discussed in Section \ref{section:expansion}, our formulation discretizes the $\chi$-dependence of quantities as exponential Fourier series. For simplicity of notation, in all subsections of Appendix \ref{appendix:algebra}, we will denote functions with $\chi$ dependence by vectors storing these Fourier components. For example, we write 
\begin{equation*}
    X_n(\chi, \phi)= \sum_{m=-n, even/odd}^n X_{n, m}(\phi)e^{im\chi}
\end{equation*}
as 
\begin{equation*}
    X_n = \begin{bmatrix}
        X_{n, -n}(\phi) \\
        X_{n, -n+2}(\phi) \\
        \vdots \\
        \text{$n+1$ components}\\
        \vdots \\
        X_{n, n}(\phi)
    \end{bmatrix}
\end{equation*}
In Appendix \ref{appendix:algebra}, we will frequently refer to the left or right-hand side of the ordered equation set, $\{J_n, \textbf{C}_n, \mathrm{I}_n, \mathrm{II}_n, \mathrm{III}_n\}$. The sides of the ordered equations will be denoted as part of the superscript, like $J_n^\text{lhs}$. Often, we will need to collect all terms that do not contain a given variable, such as $X_n$, from an ordered expression, such as $(J^\text{rhs}_n-J^\text{lhs}_n)$. This will be denoted as:
\begin{equation*}
    (J^\text{rhs}_n-J^\text{lhs}_n)|_{X_n=0}.
\end{equation*}

\subsection{Governing equations}\label{appendi:govern}
This appendix contains the full expressions for the governing equations. Substituting $\mathbf{x}$ into the Jacobian equation, \eqref{governing:J vec}, yields:
\begin{equation}\label{governing:J}\tag{$J$}
\begin{split}
    \frac{B_\alpha^2}{B^2} = B_\alpha J = & (\nabla\psi\times\nabla\theta\cdot\nabla\phi)^{-1}\\
    &=\left(
        \bar{\iota}\partial_\chi X + \partial_\phi X + \frac{dl}{d\phi}\tau Y + \frac{dl}{d\phi}\kappa Z
    \right)^2\\
    +&\left(
        \bar{\iota}\partial_\chi Y + \partial_\phi Y - \frac{dl}{d\phi}\tau X 
    \right)^2\\
    +&\left(
        \bar{\iota}\partial_\chi Z + \partial_\phi Z - \frac{dl}{d\phi}\kappa X  + \frac{dl}{d\phi} 
    \right)^2.\\
\end{split}
\end{equation}

The Frenet components of \eqref{governing:C vec}, after substituting $\mathbf{x}$, are:
\begin{equation} \label{governing:Cb}\tag{$C_b$}
    \begin{split}
        &-(B_\alpha-\bar{\iota}B_\theta)(\partial_\chi X \partial_\psi Y -\partial_\psi X \partial_\chi Y)\\
        &-B_\psi\left[
            \left(
                \partial_\phi X
                +\frac{dl}{d\phi}\tau Y
                +\frac{dl}{d\phi}\kappa Z
            \right)\partial_\chi Y 
            -\left(
                \partial_\phi Y
                -\frac{dl}{d\phi}\tau X
            \right)\partial_\chi X 
        \right]\\
        &+B_\theta\left[
             \left(
                \partial_\phi X
                +\frac{dl}{d\phi}\tau Y
                +\frac{dl}{d\phi}\kappa Z
            \right)\partial_\psi Y
            -\left(
                \partial_\phi Y
                -\frac{dl}{d\phi}\tau X
            \right)\partial_\psi X 
        \right]\\
        &=\left(\partial_\phi Z - \frac{dl}{d\phi}\kappa X + \frac{dl}{d\phi} \right)+\bar{\iota}\partial_\chi Z,
    \end{split}
\end{equation}
\begin{equation} \label{governing:Ckappa}\tag{$C_\kappa$}
    \begin{split}
        &-(B_\alpha-\bar{\iota}B_\theta)(\partial_\chi Y \partial_\psi Z -\partial_\psi Y \partial_\chi Z)\\
        &-B_\psi\left[
            \left(
                \partial_\phi Y
                -\frac{dl}{d\phi}\tau X
            \right)\partial_\chi Z 
            -\left(
                \partial_\phi Z
                -\frac{dl}{d\phi}\kappa X 
                +\frac{dl}{d\phi}
            \right)\partial_\chi Y 
        \right]\\
        &+B_\theta\left[
             \left(
                \partial_\phi Y
                -\frac{dl}{d\phi}\tau X
            \right)\partial_\psi Z
            -\left(
                \partial_\phi Z
                -\frac{dl}{d\phi}\kappa X 
                +\frac{dl}{d\phi}
            \right)\partial_\psi Y 
        \right]\\
        &=\left(\partial_\phi X + \frac{dl}{d\phi}\tau Y + \frac{dl}{d\phi}\kappa Z \right)+\bar{\iota}\partial_\chi X,
    \end{split}
\end{equation}
\begin{equation} \label{governing:Ctau}\tag{$C_\tau$}
    \begin{split}
        &-(B_\alpha-\bar{\iota}B_\theta)(\partial_\chi Z \partial_\psi X -\partial_\psi Z \partial_\chi X)\\
        &-B_\psi\left[
            \left(
                \partial_\phi Z
                -\frac{dl}{d\phi}\kappa X 
                +\frac{dl}{d\phi}
            \right)\partial_\chi X 
            -\left(
                \partial_\phi X
                +\frac{dl}{d\phi}\tau Y
                +\frac{dl}{d\phi}\kappa Z
            \right)\partial_\chi Z 
        \right]\\
        &+B_\theta\left[
             \left(
                \partial_\phi Z
                -\frac{dl}{d\phi}\kappa X 
                +\frac{dl}{d\phi}
            \right)\partial_\psi X
            -\left(
                \partial_\phi X
                +\frac{dl}{d\phi}\tau Y
                +\frac{dl}{d\phi}\kappa Z
            \right)\partial_\psi Z 
        \right]\\
        &=\left(\partial_\phi Y - \frac{dl}{d\phi}\tau X\right)+\bar{\iota}\partial_\chi Y,
    \end{split}
\end{equation}
We also use its projections along $\frac{\partial \textbf{x}}{\partial \psi}$ and $\frac{\partial \textbf{x}}{\partial \chi}$ to solve for $Y_{n,1}$ and $B_\psi$:
\begin{equation}\label{governing:Cdxdpsi}\tag{$C^{\partial\textbf{x}/\partial\psi}$}
    \begin{split}
        B_\psi J = 
        &\partial_\psi X\left(
            \bar{\iota}\partial_\chi X + \partial_\phi X + \frac{dl}{d\phi}\tau Y + \frac{dl}{d\phi}\kappa Z
        \right)\\
        +&\partial_\psi Y\left(
            \bar{\iota}\partial_\chi Y + \partial_\phi Y - \frac{dl}{d\phi}\tau X 
        \right)\\
        +&\partial_\psi Z\left(
            \bar{\iota}\partial_\chi Z + \partial_\phi Z - \frac{dl}{d\phi}\kappa X  + \frac{dl}{d\phi} 
        \right).
    \end{split}
\end{equation}
\begin{equation}\label{governing:Cdxdchi}\tag{$C^{\partial\textbf{x}/\partial\chi}$}
    \begin{split}
        B_\theta J = 
        &\partial_\psi X\left(
            \bar{\iota}\partial_\chi X + \partial_\phi X + \frac{dl}{d\chi}\tau Y + \frac{dl}{d\phi}\kappa Z
        \right)\\
        +&\partial_\psi Y\left(
            \bar{\iota}\partial_\chi Y + \partial_\phi Y - \frac{dl}{d\chi}\tau X 
        \right)\\
        +&\partial_\psi Z\left(
            \bar{\iota}\partial_\chi Z + \partial_\phi Z - \frac{dl}{d\chi}\kappa X  + \frac{dl}{d\phi} 
        \right).
    \end{split}\
\end{equation}
The projections of the force balance equation \eqref{eq:original MHS} are:
\begin{subequations}\label{governing:MHD overall}
    \begin{equation*}\label{governing:MHD I}\tag{$\mathrm{I}$}
        \frac{(\partial_\phi+\bar{\iota}\partial_\chi)\Delta}{B^2} + \frac{(\partial_\phi+\bar{\iota}\partial_\chi)p_\perp}{B^4}=\frac{\bar{\iota}}{2}\Delta\partial_\chi\left(\frac{1}{B^2}\right),
    \end{equation*}
    \begin{equation*}\label{governing:MHD II}\tag{$\mathrm{II}$}
        \frac{A_\alpha(1-\Delta)}{B^2} =
        -\frac{B_\theta}{B^4}\partial_\phi p_\perp 
        +\frac{B_\alpha-\bar{\iota}B_\theta}{B^4}\partial_\chi p_\perp 
        -\frac{B_\alpha-\bar{\iota}B_\theta}{2}\Delta\partial_\chi \left(\frac{1}{B^2}\right),
    \end{equation*}
    \begin{equation*}\label{governing:MHD III}\tag{$\mathrm{III}$}
        \frac{A_\psi(1-\Delta)}{B^2} 
        +\frac{B_\psi(\partial_\phi+\bar{\iota}\partial_\chi)\Delta}{B^2} 
        -\frac{1}{2}B_\alpha\Delta\partial_\psi\left(\frac{1}{B^2}\right)
        +\frac{J\partial_\psi p_\perp}{B^2}=0,
    \end{equation*}
\end{subequations}
where $A_\alpha \equiv (\partial_\phi+\bar{\iota}\partial_\chi)B_\theta$ and $A_\psi \equiv \partial_\psi B_\alpha - (\partial_\phi+\bar{\iota}\partial_\chi)B_\psi-B_\theta\partial_\psi\bar{\iota}$.

\subsection{Linear operators}\label{appendix:linear}
This appendix summarizes the implementation of some basic arithmetic operations, $\chi$-derivatives and $\chi$-anti-derivatives. As in Sec.\ref{section:expansion:expansion of quantities}, \texttt{pyAQSC} represents coefficients of $\epsilon^n$ that are functions of $\chi$ or $\{\chi, \phi\}$ as finite, even/odd Fourier series in $\chi$ with maximum mode number $n$. Denote such a coefficient as a vector:
\begin{equation*}
    X(\phi)=
    \begin{bmatrix}
        X_{-n}(\phi) \\
        X_{-n+2}(\phi) \\
        \vdots \\
        \text{$n+1$ components}\\
        \vdots \\
        X_{n}(\phi)
    \end{bmatrix}
    \equiv \sum_{m=-n, even/odd}^n X_{m}(\phi)e^{im\chi}
\end{equation*}
Since $\phi$-dependence is stored on grid points, all algebraic operations between such vectors and matrix elements can be performed element-wise. For simplicity, we omit all $\phi$ dependence in this appendix.

Sum and subtraction of two functions of $\chi$ are performed element-wise. Only sums and subtractions between vectors with the same even/oddness are permitted.

Scalar multiplication is represented by the discrete convolution of Fourier coefficients. For vector $A$ and $B$ with highest mode number $a$ and $b$: 
\begin{equation}\label{eq: appendix conv}
    \begin{split}
    &A(\chi)B(\chi) \\
    &= \hat{\textbf{C}}_A B_{b}\\
    &\equiv \begin{bmatrix}
        &A_{-a} &0 &0 &... &0 \\
        &A_{-a+2} &A_{-a} &0 &... &0  \\
        &A_{-a+4} &A_{-a+2} &A_{-a} &... &0  \\
        &\vdots &\vdots &\vdots &\begin{split}
            (a+b&+1)\times(b+1)\\
            &\text{total elements}
        \end{split}&\vdots  \\
        &0 &0 &0 &... &A_{a}  \\
    \end{bmatrix}
    \textbf{B}_{b},
    \end{split}
\end{equation}
where $\hat{\textbf{C}}_A$ is the convolution matrix with kernel $\vec{A}$. Conversely, when a finite, even/odd Fourier series $B(\chi)$ satisfying $A(\chi)B(\chi)=C(\chi)$ is known to exist, the scalar division $C(\chi)/A(\chi)$ is equivalent to the linear deconvolution problem:
$$
B = "\hat{\textbf{C}}_A^{-1}"C.
$$
Fortunately, all algebraic divisions we will encounter in the expansion have divisors $\vec{A}_{0}$ with only one component. In this case, the deconvolution is trivial and can be performed element-wise.

The derivative $\frac{\partial}{\partial\chi}$ is performed by the diagonal matrix:
\begin{equation*}
    \hat{\textbf{D}}_{\chi}(n) \equiv \begin{bmatrix}
        &-ni &0  &... &0 \\
        &0 &-(n-2)i  &... &0  \\
        &\vdots &\vdots &\begin{aligned}
            \text{(a zero }&\text{center }\\
            \text{element }&\text{for even $n$)}
        \end{aligned} &\vdots \\
        &0 &0 &... &ni  \\
    \end{bmatrix}.
\end{equation*}
The anti-derivative $\int{d\chi}$ is performed by the diagonal matrix:
\begin{equation*}
    \hat{\textbf{A}}_{\chi}(n) \equiv \begin{bmatrix}
        &i/n &0  &... &0 \\
        &0 &i/(n-2)  &... &0  \\
        &\vdots &\vdots &\begin{aligned}
            \text{(a zero }&\text{center }\\
            \text{element }&\text{for even $n$)}
        \end{aligned} &\vdots \\
        &0 &0 &... &-i/n  \\
    \end{bmatrix}.
\end{equation*}
When solving for $Y_n$, we encounter the under-determined linear system \eqref{recursion:Y linear}:
$$
\hat{\textbf{L}}(n)Y_n = \barB_{n+1},
$$
where $Y_n$ is a vector with $(n+1)$ components, and $\hat{\textbf{L}}(n)$ is a rank-$n$, $(n+2)\times(n+1)$ matrix, and $\barB_{n+1}$ is an $(n+2)$-component vector with $n$ linearly independent components. This system can be solved to a free component in $Y_n$. Choosing $Y_{n,0}$ or $Y_{n,1}$ as the free component at even or odd orders. Remove the corresponding column, and the first and last rows from $\hat{\textbf{L}}$:
\begin{equation*}
    \hat{\textbf{L}}' \equiv \begin{bmatrix}
        &l_{-n, -n-1}  &... &l_{-n, m_{free}-2} &l_{-n, m_{free}+2} &... &l_{-n-2, n+1} \\
        &\vdots &\ddots  &\vdots &\vdots &\ddots  &\vdots \\
        &l_{n, -n-1}  &... &l_{n, m_{free}-2} &l_{n, m_{free}+2} &... &l_{n, n+1} \\
    \end{bmatrix},
\end{equation*}
where $l_{n,m}$ are the components of $\hat{\textbf{L}}$, and $m_{free}$ is 0 or 1. $\hat{\textbf{L}}'$ is now an $n\times n$ invertible matrix for all non-trivial $B_{\alpha0}$. The solution to the system, then, is:
$$
Y_{n} = \bar{\hat{\textbf{L}}}(n)b + Y_{n,m_{free}}K_{m_{free}}(n),
$$
Where $\bar{\hat{\textbf{L}}}(n)$ is a $(n+1)\times(n+2)$ matrix defined by zero-padding the inverse of $\hat{\textbf{L}}'$ to $(n)\times(n+2)$, and then adding a zero row above the row $m_{free}$. Denote the elements of $(\hat{\textbf{L}}')^{-1}$ as $l'^{-1}$, then:
\begin{equation}\label{eq: appendix L partial inv}
    \bar{\hat{\textbf{L}}} \equiv
    \begin{bmatrix}
        &0 &l'^{-1}_{-n-1, -n-1}  
        &... &l'^{-1}_{-n-1, n+1} &0 \\
        &0 &l'^{-1}_{-m_{free}-2, -n-1}  
        &... &l'^{-1}_{-m_{free}-2, n+1} &0 \\
        &0  &0 &... &0 &0 \\
        &0 &l'^{-1}_{-m_{free}, -n-1}  
        &... &l'^{-1}_{-m_{free}, n+1} &0 \\
        &0 &l'^{-1}_{n+1, -n-1}  
        &... &l'^{-1}_{n+1, n+1} &0 
    \end{bmatrix}
\end{equation}
and the vector $K_{m_{free}}(n)$ is defined by applying $\bar{\hat{\textbf{L}}}$ to the $m_{free}$ column of $\hat{\textbf{L}}$, then setting the $m_{free}$ element to one:
\begin{equation}\label{eq: appendix C free}
    \begin{split}   
        K_{m_{free}}&\equiv
        -\bar{\hat{\textbf{L}}} L_{m_{free}}
        (\text{the } m_{free} \text{-th component of the product is 0})\\
        &+
        \begin{bmatrix}
            &0 &\\
            &\vdots &\\
            &1 &(m_{free}\text{-th component})\\
            &\vdots\\
            &0\\
        \end{bmatrix}
    \end{split}
\end{equation}

When solving for $\Delta_n$, we encounter the first-order linear PDE:
$$
(\frac{\partial}{\partial\chi} + \bar\iota_0\frac{\partial}{\partial\phi})\Delta_n(\chi, \phi) = F_n(\chi, \phi).
$$
This equation is equivalent to $n+1$ first-order linear ODEs:
\begin{equation}
    (im + \bar\iota_0\frac{\partial}{\partial\phi})\Delta_{n,m}(\phi) = F_{n,m}(\phi).
\end{equation}
At even orders, the $m=0$ component is a special case:
$$
a\frac{\partial}{\partial\phi}\Delta_{n,0}(\phi) = F_{n,0}(\phi).
$$
Here, the integration constant $\bar\Delta_{n,0}\equiv\int_0^{2\pi}d\phi\Delta_{n0}$ is left free. Note that in a trigonometric Fourier representation, the differential operator is anti-diagonal, and the PDE is equivalent to $(n+1)/2$ pairs of coupled ODEs:
$$
\pm m \Delta^{cos/sin}_{n,m}(\phi) + \bar\iota_0\frac{\partial}{\partial\phi}\Delta^{si n/cos}_{n,m}(\phi) = F^{sin/cos}_{n,m}(\phi).
$$
Compared to the exponential representation, this is trickier to implement.

\subsection{Magnetic recursion relations}\label{appendix:recursion:magnetic}

This appendix lists the explicit formulae for all recursion relations required to solve the magnetic equations. 

\paragraph{\underline{Expression for $B_{\psi n-2}$}}
$B_\psi$ is obtained by eliminating $Z_n$ from \eqref{governing:Cdxdchi} and \eqref{governing:Cdxdpsi}:
\begin{equation}\label{recursion:B_psi}
    \begin{split}
        \hat{\textbf{D}}_{\chi}(n) B_{\psi n-2}
        &=\left.
            \frac{
                \frac{\partial}{\partial\chi}(C_{\partial\textbf{x}/\partial\psi,n-2}^{\text{rhs}} - C_{\partial\textbf{x}/\partial\psi,n-2}^{\text{lhs}})
            }{B_{\alpha 0}\barB_0}
        \right|_{B_{\psi n-2}=0}\\
        &+\left.
            \frac{
                \frac{n}{2}(C_{\partial\textbf{x}/\partial\chi,n}^\text{lhs} - C_{\partial\textbf{x}/\partial\chi,n}^\text{rhs})
            }{B_{\alpha 0}\barB_0}
        \right|_{B_{\psi n-2}=0}
    \end{split}
\end{equation}
which uses $\{B_{\theta n}, 
        X_{n-1}, Y_{n-1}, Z_{n-1}, B_{\psi n-3}, \bar\iota_{\frac{n-2}{2} \text{ or } \frac{n-3}{2} }, \barB_{n-2}, B_{\alpha \frac{n-2}{2} \text{ or } \frac{n-3}{2}}\}$ and lower orders. At odd $n$, \eqref{recursion:B_psi} gives the full value of $B_{\psi n-2}$. At even $n$, the $m=0$ component of the equation is a special case where the $B_{\psi n-2,0}$ term drops out. In this case,   \eqref{recursion:B_psi} gives $B_{\psi n-2, m>0}$, and $B_{\psi n-2, 0}$ remains free. Note that unlike in \eqref{recursion:Y}, the remaining components of $B_{\psi n-2}$ do not depend on $B_{\psi n-2,0}$. 

\paragraph{\underline{Expression for $Z_n$}}
\begin{equation}\label{recursion:Z}
    Z_n=\left.\frac{(C_{\tau,n-1}^\text{rhs} Y_1 + C_{\kappa,n-1}^\text{rhs} X_1) - (C_{\tau,n-1}^\text{lhs} Y_1 + C_{\kappa,n-1}^\text{lhs} X_1)}{-ndl/d\phi}\right|_{Z_n=0}.
\end{equation}
using $\{B_{\psi n-2}, X_{n-1}, Y_{n-1}, B_{\theta n-1}, B_{\alpha \frac{n-2}{2} \text{ or } \frac{n-3}{2}},\bar\iota_{\frac{n-2}{2} \text{ or } \frac{n-3}{2}}\}$ and lower orders. 

The $m=n+1$ components of  $C_{\tau, n-1} Y_1 + C_{\kappa, n-1} X_1$ contain only contributions from $B_{\theta n,\pm n}$, indicating that $B_{\theta n,\pm n}=0$.

\paragraph{\underline{Expression for $X_n$}}
$X_n$ is given simply by
\begin{equation}\label{recursion:X}
    X_n=\left.\frac{J^\text{rhs}_n-J^\text{lhs}_n}{2\kappa(dl/d\phi)^2}\right|_{X_n=0}\\
\end{equation}
using $\{Z_{n-1}, X_{n-1}, Y_{n-1}, B^-_{ n}, B_{\alpha \frac{n}{2} \text{ or } \frac{n-1}{2}},\bar\iota_{\frac{n-2}{2}\text{ or }\frac{n-3}{2}}\} 
$ and lower orders.

\paragraph{\underline{Expression for $Y_n$}}
$Y_n$ is given by the linear, first-order ODE in $\chi$
\begin{equation}\label{recursion:Y recursion}
    \left[
            -\frac{1}{2}
            B_{\alpha 0}\frac{\partial X_1}{\partial \chi}
            +\frac{1}{2}
            B_{\alpha 0}X_1 
            \frac{\partial}{\partial \chi}
        \right]Y_n = (C_{b,n-1}^\text{rhs} - C_{b,n-1}^\text{lhs})|_{Y_n=0},
\end{equation}
which uses $\{X_n,  Y_{n-1}, Z_{n-1}, B_{\psi n-3}, B_{\theta n-1}, \bar\iota_{\frac{n-4}{2} \text{ or } \frac{n-3}{2}}, B_{\alpha \frac{n-2}{2} \text{ or } \frac{n-1}{2}}\}$ and lower orders. The ODE always has a unique periodic solution for all $B_{\alpha0}, X_1,(C_{b,\text{rhs}}^{n-1} - C_{b,\text{lhs}}^{n-1})\neq0$. To solve the equation, we write it as a linear system
\begin{equation}\label{recursion:Y linear}
    \hat{\textbf{L}}(n)Y_n = (C_{b,n-1}^\text{rhs} - C_{b,n-1}^\text{lhs})|_{Y_n=0},
\end{equation}
where $\hat{\textbf{L}}$ is an $(n+2)\times(n+1)$  operator acting on the $\chi$-Fourier coefficients of $Y_n$:
\begin{equation}\label{recursion:Y diff op}
    \begin{split}
        \hat{\textbf{L}}(n)Y_n 
        &= \left[
            -\frac{1}{2}\hat{\textbf{C}}_{
            B_{\alpha 0}(\partial X_1/\partial \chi)
            }
            +\frac{1}{2}\hat{\textbf{C}}_{
            B_{\alpha 0}X_1 
            } \hat{\textbf{D}}_{\chi}(n)
        \right]Y_n\\
        &=\left(-\frac{n}{2}B_{\alpha 0}\frac{\partial X_1}{\partial\chi} + \frac{B_{\alpha 0}X_1}{2}\frac{\partial}{\partial\chi}\right)Y_n.
    \end{split}
\end{equation}
Here,  $\hat{\textbf{C}}_{B_{\alpha 0}(\partial X_1/\partial\chi)}$ and $\hat{\textbf{C}}_{B_{\alpha 0}X_1}$ are $(n+2)\times(n+1)$ convolution matrices using $B_{\alpha 0}(\partial X_1/\partial\chi)$ and $B_{\alpha 0}X_1$ as kernels. $\hat{\textbf{D}}_{\chi}(n)$ is a $(n+1)\times(n+1)$ diagonal $\chi$ differential operator defined in \ref{appendix:linear}. 

Counting in \cite{rodriguez_magnetic} and \cite{rodriguez_mhd_1} shows that the RHS of \eqref{recursion:Y linear} contains only $n$ linearly independent components, indicating \eqref{recursion:Y linear} is under-determined with one degree of freedom. Choosing $Y_{n,0}$ or $Y_{n,1}$ as the free component, the solution to \eqref{recursion:Y linear} is:
\begin{equation}\label{recursion:Y}
    \begin{split}
        Y_{\text{odd }n} &= \bar{\hat{\textbf{L}}}(n)
            (C_{b,n-1}^\text{rhs} - C_{b,n-1}^\text{lhs})|_{Y_n=0} 
            + Y_{n,1}K_m(n)
        \\
        Y_{\text{even }n} &= \bar{\hat{\textbf{L}}}(n)
            (C_{b,n-1}^\text{rhs} - C_{b,n-1}^\text{lhs})|_{Y_n=0} 
            + Y_{n,0}K_m(n),
    \end{split}
\end{equation}
where $\bar{\hat{\textbf{L}}}$ and $K_m$ are derived in \ref{appendix:linear}. This equation solves for all remaining components of $Y_n$  using $\{Y_{n,0}$ or $Y_{n,1}, B_{\theta n-1}, 
        X_{n}, Y_{n-1}, Z_{n-1}, B_{\psi n-3}, \bar\iota_{\frac{n-2}{2} \text{ or } \frac{n-3}{2} }, B_{\alpha \frac{n}{2} \text{ or } \frac{n-1}{2}}\}
$ and lower orders.

At odd $n$, the $m=0$ component  $\eqref{recursion:B_psi}_{n+1}$ does not depend on $B_{\psi n-1,0}$. Can we use this component to further constrain $Y_{n, 1}$? Upon further inspection, we notice that the $Z_{n+1}$ dependence in $\left.\eqref{governing:Cdxdchi}_{n+1}\right|_{m=0}$ conveniently cancels out. This allows us to substitute \eqref{recursion:Y linear} into $\left.\eqref{governing:Cdxdchi}_{n+1}\right|_{m=0}$, and obtain:
\begin{equation}\label{recursion:Y ODE}
   \begin{split}
       &\left\{\left[
            \frac{dl}{d\phi}\tau\frac{\partial X_1}{\partial\chi}
            +\left(
                -\frac{dl}{d\phi}\tau X_1
                +\frac{\partial Y_1}{\partial \phi}
                +2\bar{\iota}_0
                +\frac{\partial Y_1}{\partial \chi} 
            \right)\frac{\partial}{\partial\chi}
            +\frac{\partial Y_1}{\partial\chi} \frac{\partial}{\partial\phi} 
        \right] [K_1(n)Y_{n,1}]\right\}_{m=0} \\
        &=(C_{\partial\textbf{x}/\partial\chi, n+1}^\text{lhs}-           
        C_{\partial\textbf{x}/\partial\chi, n+1}^\text{rhs})_{m=0}|_{Y_{n,1}=0}.
   \end{split}
\end{equation}
This ODE evaluates $Y_{n, 1}$ using $\{B_{\theta n+1}, 
        X_{n}, Y_{n-1}, Z_{n},\bar\iota_{\frac{n-1}{2}}, \barB_{n-1}, B_{\alpha \frac{n-1}{2}}\}
$. As the equation has no unique solution, an initial condition, $\bar{Y}_{n,1}^c \equiv \frac{1}{2\pi}\int_0^{2\pi}d\phi Y_{n,1}^c$ must also be provided.

\paragraph{\underline{Special considerations for even order $Y$}}
\par
As discussed in \cite{rodriguez_magnetic} and \cite{rodriguez_mhd_1}, the periodicity of \eqref{recursion:Y ODE} constrains one of three free parameters, initial tilt $\sigma_n(0)\equiv\left.\frac{Y^c_{n1}}{Y^s_{n1}}\right|_{\phi=0}$, the rotational transform $\bar\iota_{(n-1)/2}$, or the toroidal average of $B_{\theta n}$, $\bar{B}_{\theta n,0}\equiv \frac{1}{2\pi}\int_0^{2\pi}d\phi B_{\theta n,0}$, given the other two. In this paper, for simplicity, we treat the initial tilt $\sigma_n(0)$ as the unknown. \eqref{recursion:Y ODE} is solved with a spectral method, as Sec.\ref{section:numerical} and App.\ref{appendix:spectral} will discuss.  The nature of this equation is altered when force balance is considered. This will be discussed in greater detail in \ref{section:governing:force balance} and \ref{appendix:looped}.

    \paragraph{\underline{Input counting}}
The required inputs are 
\begin{gather*}
    \{B_{\theta n,m}(\phi) \text{ (includes $\bar{B}_{\theta n,0}$)}, \\
    B_{\psi n-2,0}(\phi), Y_{n,0}(\phi),\bar{Y}^c_{n,1}\\
    \barB_{n,m}, B_{\alpha(n-1)/2\text{ or }(n-2)/2}, \\
    \bar\iota_{(n-1)/2\text{ or }(n-2)/2}, \\
    R(\Phi), Z(\Phi) \},
\end{gather*}
where $R(\Phi)$, $Z(\Phi)$ parameterize the axis in the cylindrical coordinate $\{R, \Phi, Z\}$. At order $n$, this consists of 
$$
\left[1+2+...+(n-1)\right]+\lfloor\frac{n}{2}\rfloor+\left(\lfloor\frac{n}{2}\rfloor+1\right) + 2
=\frac{(n-1)n}{2}+2\lfloor\frac{n}{2}\rfloor+3
$$
periodic 1D functions, and
$$
\left[1+2+...+(n+1)\right]+\lfloor\frac{n+1}{2}\rfloor+\lfloor\frac{n+1}{2}\rfloor
=\frac{(n+1)(n+2)}{2}+2\lfloor\frac{n+1}{2}\rfloor
$$
scalars.

\subsection{MHS recursion relations}\label{appendix:recursion:MHD}

This appendix lists the explicit formulae for all recursion relations derived from force balance equations that can be evaluated as algebraic expressions. As discussed in \ref{section:governing:force balance} , the recursion relations for the MHS iteration are highly coupled. The coupled equations form a linear PDE system called the "looped equations". For the derivation and expression of the looped equations, see Appendix \ref{appendix:looped}.

\paragraph{\underline{Expression for $p_{\perp n}$}}
$p_{\perp n}$ is given by
\begin{equation}\label{recursion:p}
    p_{\perp n}=\left.\frac{2(E6^\text{rhs}_{n-2} - E6^\text{lhs}_{n-2})}{nB_{\alpha 0}(\barB_0)^2}\right|_{p_{\perp n}=0},
\end{equation}
using $\{B_{\psi n-2}, B_{\theta n-2}, \Delta_{n-1},\bar\iota_{\frac{n-2}{2} \text{ or } \frac{n-3}{2} }, \barB_{n}, B_{\alpha \frac{n}{2} \text{ or } \frac{n-1}{2}} \}$ and lower orders. Here \eqref{governing:MHD E6} is a transformation of \eqref{governing:MHD III}:
\begin{equation}\label{governing:MHD E6} \tag{$E6$}
\begin{split}
    \frac{B_\alpha}{B^2}\partial_\psi p_\perp &= \left(\frac{\partial}{\partial\phi}+ \bar{\iota} \frac{\partial}{\partial\chi}\right)[B_\psi(1-\Delta)]\\
    &+(\Delta-1)(\frac{\partial B_\alpha}{\partial\psi} -B_\theta \frac{\partial \bar\iota}{\partial\psi}) + \frac{1}{2}B^2B_\alpha\Delta\frac{\partial}{\partial\psi}\left(\frac{1}{B^2}\right)\Leftrightarrow\eqref{governing:MHD III}.
\end{split} 
\end{equation}
\paragraph{\underline{Expression for $\Delta_n$}}
$\Delta_n$ is given by \eqref{governing:MHD I}:
$$
\left(\frac{\partial}{\partial\phi}+ \bar{\iota}_0 \frac{\partial}{\partial\chi}\right)\Delta_n = \left.\frac{I^\text{rhs}_n - I^\text{lhs}_n}{\barB_0}\right|_{\Delta_n=0},
$$
using $\{p_{\perp n}, \Delta_{n-1}, \bar\iota_{\frac{n-2}{2} \text{ or } \frac{n-1}{2} }, \barB_{n}\}$ and lower orders. The equation's $\chi$ Fourier coefficients correspond to $n+1$ inhomogeneous linear, first-order ODEs:
\begin{equation}\label{recursion:Delta}
    \left(\frac{\partial}{\partial\phi}+ im\bar{\iota}_0 \right)\Delta_{n,m} = \left.\frac{I^\text{rhs}_{n,m} - I^\text{lhs}_{n,m}}{\barB_0}\right|_{\Delta_n=0}.
\end{equation}
At odd $n$, \eqref{recursion:Delta} gives the full value of $\Delta_{n,m}$. At even $n$, \eqref{recursion:Delta} gives the full value of $\Delta_{n,m\neq0}$, and $\Delta_{n,m=0}$ to an integration constant $\bar{\Delta}_{n,0}=\frac{1}{2\pi}\int_0^{2\pi}\Delta_{n,0}d\phi$.

\paragraph{\underline{The looped equations: $B_{\psi n,0}$, $Y^\text{free}_{n}$, $B_{\theta n}$ and $\bar{\Delta}_{n,0}$}}
\par
The remaining unknowns, $B_{\psi n,0}$, $Y^\text{free}_{n}$, $B_{\theta n}$ and $\bar{\Delta}_{n,0}$ are solved self-consistently from a system of coupled inhomogeneous, linear, 3rd-order ODEs constructed from \eqref{governing:MHD II}, \eqref{recursion:Y ODE}, \eqref{recursion:p} and \eqref{recursion:B_psi}. We refer to this system as the looped equations. The construction of the looped equations is detailed in App.\ref{appendix:looped}.

\paragraph{\underline{Input counting}}
The required inputs are: 
\begin{gather*}
    \{\bar{B}_{\theta n,0}, \barB_{n,m}, p_{\perp0,0}(\phi), \bar\Delta_0, \\
    B_{\alpha(n-1)/2\text{ or }(n-2)/2},\\
    \bar\iota_{(n-1)/2\text{ or }(n-2)/2},\\
    R(\Phi), Z(\Phi)\}.
\end{gather*}
To order $n$, this includes 3 periodic 1D functions and 
\begin{equation*}
\begin{split}
    &1+1+\left[1+2+...+(n+1)\right]+\lfloor\frac{n+1}{2}\rfloor+\lfloor\frac{n+1}{2}\rfloor \\
=&2+\frac{(n+1)(n+2)}{2}+2\lfloor\frac{n+1}{2}\rfloor \text{ free scalars.}
\end{split}
\end{equation*}

\subsection{The looped equations}\label{appendix:looped}
\cite{rodriguez_mhd_1} shows that a system of coupled, linear, 3rd order ODEs needs to be solved to obtain $\{B_{\theta n,m}, Y_{n,0}, B_{\psi n-2,0}, \bar\Delta_{n,0}\}$ at even $n$, and $\{B_{\theta n,m}, B_{\theta n+1,0}, Y_{n,1}\}$ at odd $n$. This appendix summarizes the procedure for constructing and solving this system.

We start by eliminating all order $n+1$ terms in $\mathrm{II}_{n+1}$:
\begin{equation*}
    \begin{split}
        &B_{\alpha0} (\barB_0)^2 \partial_\chi p_{\perp n+1} \\
        - &(\barB_0)^2 (\partial_\phi p_{\perp 0}) B_{\theta n+1} \\
        + &\barB_0(\Delta_0-1) 
        (\partial_\phi + \bar\iota_0 \partial_\chi) 
        B_{\theta n+1}\\
        + &\text{rest of }(\mathrm{II}^\text{rhs}_{n+1}-\mathrm{II}^\text{lhs}_{n+1}) = 0
    \end{split}
\end{equation*}
First, substitute $\partial_\chi p_{\perp n+1}$ with $\partial_\chi\eqref{recursion:p}_{n+1}^\text{rhs}$. The resulting equation is
\begin{equation*}
    \begin{split}
        B_{\alpha0} (\barB_0)^2 \partial_\chi &\left[ 
            -\frac{2}{n+1}\frac{(\partial_\phi \Delta_0)}{B_{\alpha 0} \barB_0}
            B_{\psi n-1} \right. \\
            & -\left.\frac{2}{n+1}\frac{\Delta_0-1}{B_{\alpha 0} \barB_0}
            (\bar\iota_0\partial_\chi+\partial_\phi) B_{\psi n-1}
            +\left.\eqref{recursion:p}_{n+1}^\text{rhs}\right|_{B_{\psi n-1}=0}
        \right] \\
        & -(\barB_0)^2 (\partial_\phi p_{\perp 0}) B_{\theta n+1} \\
        & +\barB_0(\Delta_0-1) 
        (\partial_\phi + \bar\iota_0 \partial_\chi) 
        B_{\theta n+1}\\
        & +\text{rest of }(\mathrm{II}^\text{rhs}_{n+1}-\mathrm{II}^\text{lhs}_{n+1}) = 0,
    \end{split}
\end{equation*}
where all $B_{\psi n-2}$ terms have $\chi$-independent coefficients. Commute these coefficients with $\partial_\chi$:
\begin{equation*}
    \begin{split}
        \cancel{B_{\alpha 0}} (\barB_0)^{\cancel{2}} &\left[ 
             -\frac{2}{n+1}\frac{(\partial_\phi \Delta_0)}{\cancel{B_{\alpha 0} \barB_0}}
              \right. \\
            & -\left.\frac{2}{n+1}\frac{\Delta_0-1}{\cancel{B_{\alpha 0} \barB_0}}
            (\bar\iota_0\partial_\chi + \partial_\phi)
        \right](\partial_\chi B_{\psi n-1}) \\
        & + B_{\alpha0} (\barB_0)^2 \partial_\chi
        \left[\left.\eqref{recursion:p}_{n+1}^\text{rhs}\right|_{B_{\psi n-1}=0}\right]\\
        & -(\barB_0)^2 (\partial_\phi p_{\perp 0}) B_{\theta n+1} \\
        & +\barB_0(\Delta_0-1) 
        (\partial_\phi + \bar\iota_0 \partial_\chi) 
        B_{\theta n+1}\\
        & +\text{rest of }(\mathrm{II}^\text{rhs}_{n+1}-\mathrm{II}^\text{lhs}_{n+1}) = 0.
    \end{split}
\end{equation*}
Substitute all $\partial_\chi B_{\psi n-1}$ with $\eqref{recursion:B_psi}_{n+1}^\text{rhs}$. By $\mathrm{I}_0\Leftrightarrow \barB_0\partial_\phi p_{\perp 0}+\partial_\phi \Delta_0=0$, all $B_{\theta n+1}$ terms in the equation cancel:
\begin{equation*}
    \begin{split}
        \barB_0&\left[ 
            -\frac{2}{n+1}(\partial_\phi \Delta_0)
            -\frac{2}{n+1}(\Delta_0-1) (\bar\iota_0\partial_\chi + \partial_\phi)
        \right] \cdot \\
        & \cdot\left[ 
            \cancel{\frac{n+1}{2}B_{\theta n+1}}
            + \left. \eqref{recursion:B_psi}_{n+1}^\text{rhs}\right|_{B_{\theta n+1}=0} \right] \\
        & + B_{\alpha0} (\barB_0)^2 \partial_\chi
        \left[\left. \eqref{recursion:p}_{n+1}^\text{rhs}\right|_{B_{\psi n-1}=0}\right]\\
        & \cancel{-(\barB_0)^2 (\partial_\phi p_{\perp 0}) B_{\theta n+1}} \\
        & \cancel{+\barB_0(\Delta_0-1) 
        (\partial_\phi + \bar\iota_0 \partial_\chi) 
        B_{\theta n+1}}\\
        & +\text{the rest of }(\mathrm{II}^\text{rhs}_{n+1}-\mathrm{II}^\text{lhs}_{n+1}) = 0,
    \end{split}
\end{equation*}
\begin{equation}\label{governing:MHD II tilde}\tag{$\widetilde{\mathrm{II}}$}
    \begin{split}
        -\barB_0\frac{2}{n+1}&\left[ 
            \partial_\phi \Delta_0
            +(\Delta_0-1) (\bar\iota_0\partial_\chi + \partial_\phi)
        \right] 
            \left. \eqref{recursion:B_psi}_{n+1}^\text{rhs}\right|_{B_{\theta n+1}=0}
         \\
        & + B_{\alpha0} (\barB_0)^2 \partial_\chi
        \left. \eqref{recursion:p}_{n+1}^\text{rhs}\right|_{B_{\psi n-1}=0}\\
        & +\text{the rest of }(\mathrm{II}^\text{rhs}_{n+1}-\mathrm{II}^\text{lhs}_{n+1}) = 0.
    \end{split}
\end{equation}
At odd $n$, the $m=0$ component of $\mathrm{II}_{n+1}$ is a special case. Here, the $\partial_\chi p_{\perp n+1}$ term vanishes, and the remaining order $n+1$ quantity $B_{\theta n+1,0}$ will be treated as an unknown:
\begin{equation*}\tag{$\mathrm{II}_0$}
    \begin{split}
        &\cancel{B_{\alpha0} (\barB_0)^2 \partial_\chi p_{\perp n+1, 0}} \\
        - &(\barB_0)^2 (\partial_\phi p_{\perp 0}) B_{\theta n+1,0} \\
        + &\barB_0(\Delta_0-1) 
        (\partial_\phi + \cancel{\bar\iota_0 \partial_\chi}) 
        B_{\theta n+1,0}\\
        + &\text{rest of }(\mathrm{II}^\text{rhs}_{n+1,0}-\mathrm{II}^\text{lhs}_{n+1,0}) = 0.
    \end{split}
\end{equation*}
Thus far, we have eliminated all order $n+1$ terms in $\mathrm{II}_{n+1}$ to obtain $\widetilde{\mathrm{II}}_{n+1}$. In \cite{rodriguez_mhd_1}, this equation is referred to as the "looped equations", since algebraic expressions for $B_{\theta n}$ always contain order $n$ quantities that depend on $B_{\theta n}$. $\widetilde{\mathrm{II}}_{n+1}$ has $n+2$ components, among which $n$ are non-zero. The expression for $\widetilde{\mathrm{II}}_{n+1}$ can be found in supplemental data.

To construct recursion relations for $\{B_{\theta n}, B_{\theta n+1,0}(\text{odd }n), B_{\psi n,0}(\text{even }n), Y^\text{free}_{n}\}$, list all order $n$ quantities in the equation (their coefficients are available in supplemental data):
\begin{equation*}
    \begin{split}
        X_n:&X_n, 
            \partial_\chi X_n, \,
            \partial_\phi X_n, \\
            &\partial_{\chi \phi} X_n, \,
            \partial_{\phi \phi} X_n, \,
            \partial_{\chi \chi} X_n, \\
            &\partial_{\chi \chi \chi} X_n, \,
            \partial_{\chi \chi \phi} X_n, \,
            \partial_{\chi \phi \phi} X_n, \\
        Y_n: &Y_n, 
        \partial_\chi Y_n, \,
        \partial_\phi Y_n, \\
        &\partial_{\chi \phi} Y_n, \,
        \partial_{\phi \phi} Y_n, \,
        \partial_{\chi \chi} Y_n, \\
        &\partial_{\chi \chi \chi} Y_n, \,
        \partial_{\chi \chi \phi} Y_n, \,
        \partial_{\chi \phi \phi} Y_n, \\
        Z_n: &Z_n, 
        \partial_\chi Z_n, \,
        \partial_\phi Z_n, \\
        &\partial_{\chi \phi} Z_n, \,
        \partial_{\chi \chi} Z_n, \\
        B_{\theta n}: &B_{\theta n}, \,
        \partial_\chi B_{\theta n}, \,
        \partial_\phi B_{\theta n},\\
        B_{\psi n-2}: &B_{\psi n-2},\,
        \partial_\phi B_{\psi n-2},\,
        \partial_\chi B_{\psi n-2},\\
        &\partial_{\chi\phi} B_{\psi n-2},\,
        \partial_{\chi\chi} B_{\psi n-2},\\
        p_{\perp n}: &p_{\perp n},\, \partial_\chi p_{\perp n},\\
        \Delta_n: &\Delta_n,\,
        \partial_\chi \Delta_n.  
    \end{split}
\end{equation*}
Substituting the order $n$ dependencies of $\{X_n, Y_n, Z_n, B_{\psi n-2}, p_{\perp n}, \Delta_n\}$ into $\widetilde{\mathrm{II}}_{n+1}$,
\begin{equation}\label{looped: substitution}
    \begin{split}
        B_{\psi n-2} = &
        \frac{n}{2}B_{\theta n} + B_{\psi n-2,0}\,(\text{ even $n$ only}) + ...
        \eqref{recursion:B_psi},\\
        Z_n = &
        (X_1\partial_\chi Y_1 - Y_1\partial_\chi X_1)B_{\psi n-2} + ...
        \eqref{recursion:Z},\\
        p_{\perp n} = 
            &-\left(\frac{2}{n}\frac{\partial_\phi\Delta_0}{B_{\alpha 0}\barB_0}\right)B_\psi\\
            &-\left(\frac{2}{n}\frac{\Delta_0-1}{B_{\alpha 0}\barB_0}\right)\partial_\phi B_\psi \\
            &-\left(\frac{2}{n}\frac{\Delta_0-1}{B_{\alpha 0}\barB_0}\bar\iota_0 \right)\partial_\chi B_\psi + ...
        \eqref{recursion:p},\\
        X_n = &
        \frac{1}{\kappa dl/d\phi}(\partial_\phi+\bar\iota_0\partial_\chi) Z_n + ...
        \eqref{recursion:X},\\
        (\partial_\phi+\bar\iota_0\partial_\chi)\Delta_n =
        &-\barB_0(\partial_\phi+\bar\iota_0\partial_\chi) p_{\perp n} + ...
        \eqref{recursion:Delta}\\
        &\Leftrightarrow (\barB_0 \text{ has no $\chi$ or $\phi$ dependence})\\
        \Delta_n = &
        -\barB_0 p_{\perp n} + ...,\\
        Y_n = & Y_{n,1}K_m(n)\\
            & + \bar{\hat{\textbf{L}}}(n)\left(\frac{n}{2}B_{\alpha 0}\partial_\chi Y_1\right) X_n\\ 
            & + \bar{\hat{\textbf{L}}}(n)\left(\frac{1}{2}B_{\alpha 0} Y_1\right) \partial_\chi X_n\\
            & + \bar{\hat{\textbf{L}}}(n)\left(...\right) \eqref{recursion:Y},
    \end{split}
\end{equation}
we can write all order $n$ terms in $\widetilde{\mathrm{II}}_{n+1}$ in terms of 
$$\{B_{\theta n}, B_{\theta n+1,0}(\text{odd }n), \partial_\phi B_{\psi n-2,0}(\text{even }n), Y^\text{free}_{n}\}$$
and lower order quantities. Applying the linear operators in \ref{appendix:linear}, its non-zero components become $n$ coupled ODEs for  
$$\{B_{\theta n}, B_{\theta n+1,0}(\text{odd }n), \partial_\phi B_{\psi n-2,0}(\text{even }n), Y^\text{free}_{n}\}.$$
The full expression for $\widetilde{\mathrm{II}}_{n+1}$ is available in the supplemental data.

At even $n$, the number of non-zero components in $\widetilde{\mathrm{II}}_{n+1}$ equals the total number of components in the unknowns, $\{B_{\theta n}, B_{\psi n-2,0}, Y_{n,0}\}$: \cite{rodriguez_mhd_1}
\begin{equation*}
    \begin{split}
         n (\text{components in }\widetilde{\mathrm{II}}_{n+1}) = n&+1 \, (B_{\theta n})\\
         &-2 \, (B_{\theta n,\pm n}=0)\\
         &-1 \, (B_{\theta n,0}\text{ is calculated at order $n-1$})\\
         &+1 \, (Y_{n,0})\\
         &+1 \, (B_{\psi n-2,0}),
    \end{split}
\end{equation*}
Here, $B_{\psi n-2,0}$ will only appear as $\partial_\phi B_{\psi n-2,0}$ and its derivatives. The periodicity of $B_{\psi n-2,0}$ constrains the free scalar parameter $\bar\Delta_{n,0}$\cite{rodriguez_mhd_2}. Substituting all differential operators with the pseudo-spectral operator, the system becomes an exactly determined linear system and can be solved by common linear algebra packages. 

At odd $n$, $\widetilde{\mathrm{II}}_{n+1}$ has one less component than the unknowns, $\{B_{\theta n}, B_{\theta n+1,0}, Y_{n,1}\}$: 
\begin{equation*}
    \begin{split}
         n (\text{components in }\widetilde{\mathrm{II}}_{n+1}) < n&+1 \\
         = n&+1 \, (B_{\theta n})\\
         &-2 \, (B_{\theta n,\pm n}=0)\\
         &+1 \, (B_{\theta n+1,0})\\
         &+1 \, (Y_{n,1}).
    \end{split}
\end{equation*}
This missing component is filled with \eqref{recursion:Y ODE},  the ODE for $Y_{n,1}$ discussed in \ref{section:recursion:magnetic}. When force balance is considered, \eqref{recursion:Y ODE} becomes a linear, inhomogeneous ODE for $\{Y_{n,1}, B_{\theta n}, B_{\theta n+1}\}$. Adding \eqref{recursion:Y ODE} to the system will now make the number of equations equal to that of the unknowns. We similarly solve the odd-order system via numerical linear algebra.

Notably, one can also treat $\bar{\iota}_{(n-1)/2}$ as an unknown and provide $\sigma_n(0)\equiv\frac{Y^c_{n,1}(\phi=0)}{Y^s_{n,1}(\phi=0)}$ or $\bar B_{\theta n+1,0}\equiv\oint d\phi B_{\theta n+1,0}$ for the equation to have a unique periodic solution. This would allow the self-consistent construction of magnetic shear $\bar\iota_1$ provided initial tilt $\sigma_1(0)$ or average toroidal current profile $\bar B_{\theta n+1,0}\equiv\oint d\phi B_{\theta 2,0}$. 

Thus far, we have constructed an ODE system solving for 
$$
\{B_{\theta n}, B_{\theta n+1,0}(\text{odd }n), B_{\psi n-2,0}(\text{even }n), Y^\text{free}_{n}\}
$$ using lower-order quantities. 

\subsection{Leading order formulae}\label{appendix:leading}
As described in \ref{section:expansion:expansion of quantities},
$$X_0=Y_0=Z_0=0.$$
First, from $\eqref{governing:J}_1$, $B_{\alpha 0}$ is not a free parameter and is related to input $\barB_0$ by:
$$
B_{\alpha 0} = \frac{dl}{d\phi}\frac{1}{\sqrt{\barB_0}}
$$
Assuming non-trivial $X_1, Y_1$, $\eqref{governing:C vec}_0, \eqref{governing:C vec}_1$ gives $B_{\theta0} = 0$.

For consistency with \cite{rodriguez_mhd_1}, and to simplify inputs, choose a coordinate where ${B}^s_{1,1}=0$. Define $\eta \equiv -\frac{B_1}{2\barB_0}$. Then, 
\begin{equation}
    B_{1,1} = B_{1,-1} = \frac{{B}^c_{1,1}}{2}
\end{equation}

For consistency with higher orders, we calculate $\Delta_0$ using $p_{\perp 0}$ and $\bar\Delta_0$ as inputs. $\eqref{governing:MHD I}_0$ relates $p_{\perp 0}$ and $\Delta_0$ to a constant parameter:
$$
\barB_0 p_{\perp 0} + \Delta_0 = const.
$$
This gives:
\begin{equation}
    \Delta_0 = -\barB_0 p_{\perp 0} + \overline{(\barB_0 p_{\perp 0})} + \bar\Delta_0,
\end{equation}
where $\overline{(\barB_0 p_{\perp 0})}$ is the $\phi$-average of $\barB_0 p_{\perp 0}$.

Then, we can find $\{X_1, p_{\perp 1}, \Delta_1\}$ from identical recursion relations to higher orders, $\{\eqref{recursion:X}_1, \eqref{recursion:p}_1, \eqref{recursion:Delta}_1\}$.

At the leading order, the equation system solving for $\{B_{\theta 2,0}, Y_1^{free}\}$ is no longer coupled. At this order, \eqref{governing:MHD II tilde} only has one component, $\eqref{governing:MHD II}_{2, 0}$. Since $B_{\theta 1}=0$, the equation becomes a 1st order linear ODE for $B_{\theta 2,0}$:
\begin{multline}\label{eq:appendix looped leading inhomog}
    -(\barB_0)^2\left(\frac{\partial}{\partial\phi}p_{\perp 0}\right) B_{\theta 2,0}
    +\barB_0(\Delta_0-1) \left(\frac{\partial}{\partial\phi} B_{\theta 2,0}\right) \\
    = -\frac{B_{\alpha 0}}{2}\left(4\barB_0B_1\frac{\partial}{\partial\chi}p_{\perp 1} - \Delta_1\frac{\partial}{\partial\chi}B_1\right)
\end{multline}

After finding $B_{\theta2,0}$, we substitute it into \eqref{recursion:Y ODE}. This gives a simple algebraic expression for $Y_{1, 1}^s$:
\begin{equation}
    Y^s_{1,1} = \frac{2\sqrt{\barB_0}}{\eta\kappa},
\end{equation}
and a Riccati equation for $Y^c_{1,1}$:

\begin{equation}\label{eq:appendix looped leading nonlin}
    (Y^c_{1,1})' = q_0 + q_1Y^c_{1,1} + q_2(Y^c_{1,1})^2,
\end{equation}
where
\begin{equation*}
    \begin{split}
        q_2 =& -\frac{\bar\iota_0\eta}{2\sqrt{\barB_0}\kappa},\\
        q_1 =& \frac{\kappa'}{\kappa}, \\
        q_0 =&- \bar\iota_0\left(\frac{2\sqrt{\barB_0}}{\eta}\kappa+\frac{\eta^3}{2\sqrt{\barB_0}}\frac{1}{\kappa^3}\right)\\
        &+\frac{dl}{d\phi}(2\tau+B_{\theta2,0})\frac{\eta}{\kappa},\\
    \end{split}
\end{equation*}
As proved in \cite{rodriguez_magnetic}, this boundary value problem has a unique solution pair $\{Y^c_{1,1}, \bar\iota_0\}$ given $\{Y_{1,1}^c(0), \eta, \bar{B}_{\theta2,0}, \kappa, \tau\}$, or  $\{Y^c_{1,1}, \bar{B}_{\theta2,0}\}$ given $\{Y_{1,1}^c(0),\eta, \bar\iota_0, \kappa, \tau\}$. In \texttt{pyAQSC}, we solve \eqref{eq:appendix looped leading nonlin} with a shooting method. For each value of $ \bar\iota_0$ or $\bar{B}_{\theta2,0}$, we can obtain the corresponding $Y_{1,1}^c(\phi)$ by numerically solving the initial value problem with the 4th order Runge-Kutta method. Treating the breaking of the boundary condition,  $f_{BC} = [Y_{1,1}^c(2\pi/n_{fp}) - Y_{1,1}^c(0)]$ as a function of $\bar\iota_0$  or  $\bar{B}_{\theta2,0}$, we can solve \eqref{eq:appendix looped leading nonlin} by performing a scalar Newton iteration:
\begin{equation}
    (\bar\iota_0 \text{ or } \bar{B}_{\theta2,0})_{n+1}=(\bar\iota_0 \text{ or } \bar{B}_{\theta2,0})_n-\frac{f_{BC}\left[(\bar\iota_0 \text{ or } \bar{B}_{\theta2,0})_n\right]}{f_{BC}^{\prime}\left[(\bar\iota_0 \text{ or } \bar{B}_{\theta2,0})_n\right]},
\end{equation}
where $f'_{BC}$ is calculated by auto-differentiation.

\section{Numerical implementations}
\subsection{Pseudo-spectral methods}\label{appendix:spectral}
\texttt{pyAQSC} stores functions of $\phi$ at $n_\phi$ uniformly spaced grid points, located at $\{\phi_m = \frac{2\pi m}{n_\phi}|m = 0, ... n_\phi-1\}$. In \texttt{JAX}, the DFT of a discrete function, $f_m = f(\phi_m)$, is: \cite{numpy}

\begin{gather*}
         F_k = \sum^{n_\phi-1}_{m=0} f_m exp\left(-2\pi i \frac{mk}{n_\phi}\right),\\
         k = \{-n_\phi/2, -n_\phi/2+1,\, ... \,,  n_\phi/2-1\} / \pi 
\end{gather*}

The discrete differential operator, $(\partial_\phi)^k_l$, is:
$$
    (\partial_\phi)^k_l = \left(- i l n_\phi\right)\delta^k_l
$$
The anti-derivative operator, $(\partial_\phi^{-1})^k_l$, is:
$$
    (\partial_\phi^{-1})^k_l = \left(\frac{i}{l n_\phi}\right)\delta^k_l
$$
Multiplication of $f_m$ with another discrete function $g_n$ is equivalent to applying a $(2n_\phi+1, n_\phi+1)$ convolution operator,
$$
(C_G)^k_l = G_{l-k},
$$
where the kernel $G_l$ is the DFT of $g_m$. This operator becomes an $(n_\phi, n_\phi)$ invertible square matrix neglecting rows of $C_G$ with $l<-n_\phi/2$ or $l>n_\phi/2-1$.

With $(\partial_\phi)^k_l$ and $(C_G)^k_l$, we can construct any discrete linear differential operator as a $(n_\phi, n_\phi)$ square matrix $L_l^k$. A linear, inhomogeneous ODE becomes a linear system:
\begin{equation*}
    \begin{split}
        (a + b\partial_\phi + c\partial_\phi^2 + ...)f(\phi_m) &= s(\phi_m),\\
        \left[(C_A)^k_l + (C_B)^j_l(\partial_\phi)^k_j+ ...\right]F_k &= S_l,\\
        L_l^kF_k&=S_l,
    \end{split}
\end{equation*}
and a linear, homogeneous ODE becomes a null-space problem:
$$
    L_l^kF_k=0
$$
We choose to remove all elements in $L_l^k$ and $S_l$ with $k,l<k_\text{cutoff}$, where $k_\text{cutoff}$ is the empirical cut-off frequency chosen at the order. Appendix \ref{appendix:low_pass} discusses our choices of $k_\text{cutoff}$ at each order. This is equivalent to applying a low-pass filter on $F_k$, and drastically reduces the time required to find a numerical solution.

When a linear inhomogeneous ODE has a unique solution under periodic boundary condition, $L^k_l$ is invertible, and $F_k$ is simply $(L^{-1})^l_kS_l$. 

When a linear inhomogeneous ODE has no unique solution(e.g., \eqref{eq:appendix looped leading inhomog}), $L^k_l$ is singular. To impose an initial condition, we add an additional row each to $L^k_l$ and $S_l$. For example, under the initial condition $\overline{f_m}=C\Leftrightarrow F_0 = n_\phi C$, the inhomogeneous ODE $L_l^kF_k=S_l$ becomes:
$$(L')^k_lF_k = S'_l,$$
$$L' = \begin{bmatrix}
        &L_{-n_\phi/2}^{-n_\phi/2} &... &L_{-n_\phi/2}^{0} &... &L_{-n_\phi/2}^{n_\phi/2} \\
        &\vdots &\ddots &\vdots &\ddots &\vdots\\
        &L_{-n_\phi/2}^{n_\phi/2} &... &L_{n_\phi/2}^{0} &... &L_{n_\phi/2}^{n_\phi/2} \\
        &0 &... &A &... &0  \\
    \end{bmatrix},$$
$$S'= \begin{bmatrix}
        &S_{-n_\phi/2} \\
        &\vdots\\
        &S_{n_\phi/2-1} \\
        &An_\phi C \\
    \end{bmatrix},$$
where $A=(\sum_{k,l}L_l^k)/n_\phi^2$ is an empirically picked scaling factor for good numerical behavior. The equation now becomes a linear least square problem, $|(L')_l^kF_k-S'_l|^2$, and can be solved with SVD. Identical procedure can be applied to solve the null space problem from a linear, homogeneous ODE as a linear least square problem. 

Both above-mentioned approaches are applicable to systems of coupled linear, inhomogeneous ODEs. Instead of vectors and matrices, we now use rank 2 tensors for $F_{ik}$ and $S_{jl}$, and a rank 4 tensor for $L^{ik}_{jl}$, where the $k,l$ indices $\phi$ mode coefficients, and $i,j$ represent $\chi$ mode coefficients discussed in \ref{appendix:linear}. After flattening $L^{ik}_{jl}$ and $S_{jl}$ into a $(n_\chi n_\phi, n_\chi n_\phi)$ matrix and a $n_\chi n_\phi$-component vector, the problem can be solved with inversion or SVD like discussed above, depending on whether additional constraints are needed.

\subsection{Symbolic order-matching}\label{appendix:cauchy}
In this appendix, we summarize the method used to obtain the symbolic expressions for $\{J_n, C_b^n, C_\kappa^n, C_\tau^n, \mathrm{I}_n,\mathrm{II}_n,\mathrm{III}^n\}$. This eliminates the need to run costly computer algebra codes at every order.

The governing equation system consists of up to 3rd order polynomials of power-Fourier series and their derivatives. Truncated at the $n$-th order, it contains $\sum_n\mathcal{O}(n^3)\approx\mathcal{O}(n^4)$ terms when the outer (power) summation is expanded. It is obviously too computationally expensive to perform order-matching by expanding the power series to the $n$-th order. Instead, we directly find a series form of the recursion relation at a given order by symbolic operations.

All terms in the expanded governing equations can be written as Cauchy products:
\begin{equation}
    \begin{split}
        F(\epsilon)\equiv\epsilon^l&\frac{d^{a_1}}{d\epsilon^{a_1}} X_1(\epsilon)\frac{d^{a_2}}{d\epsilon^{a_2}}X_2(\epsilon)...\frac{d^{a_k}}{d\epsilon^{a_k}}X_k(\epsilon)+(\text{other terms})\\
        =\epsilon^l
        &\left[\sum_{j_1=0}^{\infty}\frac{(j_1+a_1)!}{j_1!}X_{1,j_1+a_1} \epsilon^{j_1}\right]\cdot\\
        &\left[\sum_{j_2=0}^{\infty}\frac{(j_2+a_2)!}{j_2!}X_{2,j_2+a_2} \epsilon^{j_2}\right]\cdot \\
        &...\\        &\left[\sum_{j_k=0}^{\infty}\frac{(j_k+a_k)!}{j_k!}X_{k,j_k+a_k} \epsilon^{j_k}\right]+(\text{other terms}).
    \end{split}
\end{equation}
Rewrite $F(\epsilon)$ as a Cauchy product:
\begin{equation}
    \begin{split}
        F(\epsilon) = \epsilon^l
        &\sum_{j_1=0}^{\infty}\sum_{j_2=0}^{j_1}...\sum_{j_k=0}^{j_{k-1}}
        \frac{(j_1+a_1)!}{j_1!}
        \frac{(j_2+a_2)!}{j_2!}...
        \frac{(j_k+a_k)!}{j_k!}\\
        &X_{1,j_k+a_1}
        X_{2,(j_{k-1}-j_k)+a_2}...
        X_{k,(j_1-j_2)+a_k}\\
        &\epsilon^{\cancel j_k}\epsilon^{(\cancel j_{k-1}-\cancel j_k)}...\epsilon^{(j_1-\cancel j_2)}+(\text{other terms}),\\
        = &\sum_{j_1=0}^{\infty}\sum_{j_2=0}^{j_1}...\sum_{j_k=0}^{j_{k-1}}
        \frac{(j_1+a_1)!}{j_1!}
        \frac{(j_2+a_2)!}{j_2!}...
        \frac{(j_k+a_k)!}{j_k!}\\
        &X_{1,j_k+a_1}
        X_{2,(j_{k-1}-j_k)+a_2}...
        X_{k,(j_1-j_2)+a_k}\epsilon^{j_1+l}\\
        &+(\text{other terms}),
    \end{split}
\end{equation}

Simply removing the outer infinite sum and substituting all $j_1$ occurrences with $(n-l)$ yields the general formula for the nth order coefficient of the expression in a compact sum over $O(n^{k-1})$ terms:

\begin{equation}\label{eq:app cauchy}
    \begin{split}
        F_n=&\sum_{j_2=0}^{n-l}...\sum_{j_k=0}^{j_{k-1}}
        \frac{(n-l+a_1)!}{j_1!}
        \frac{(j_2+a_2)!}{j_2!}...
        \frac{(j_k+a_k)!}{j_k!}\\
        &X_{1,j_k+a_1}
        X_{2,(j_{k-1}-j_k)+a_2}...
        X_{k,(n-l-j_2)+a_k}
        +(\text{other terms})_n,
    \end{split}
\end{equation}
where
$$
F(\epsilon) = \sum_{j=0}^\infty F_j\epsilon^j.
$$

Expressions containing functions represented by skip power series slightly complicates the procedure. In these cases, the power of $\epsilon$ in the Cauchy product form is a linear function of all indices $j_1, ... j_k$. For example:

\begin{equation}
    \begin{split}
        G(\epsilon)&=\epsilon^l\frac{\partial X_{\text{diff}}(\epsilon)}{\partial\epsilon}X_{\text{skip}}(\epsilon)\\
        &=\left(\sum_{j_1=0}^{\infty}j_1 X_{\text{diff},j_1+1} \epsilon^{j_1-1+l} \right)\left( \sum_{j_2=0}^{\infty}X_{\text{skip},j_2} \epsilon^{2j_2} \right)\\
        &=\sum_{j_1=0}^{\infty}\sum_{j_2=0}^{j_1}j_2 X_{\text{diff},j_2+1} X_{\text{skip},j_1-j_2}\epsilon^{j_2+2(j_1-j_2)-1+l}.
    \end{split}
\end{equation}

The n-th order coefficient can be obtained as follows. First, solve the linear equation,
$$
j_2+2(j_1-j_2)-1+l=n \Leftrightarrow j_1 = \frac{j_2+n+1-l}{2},
$$ 
remove the infinite sum and substitute all $j_1$ occurrences with the solution:
\begin{equation}
    G_n(\epsilon) = \sum_{j_2=0}^{j_1} I(j_2) j_2 X_{\text{diff},j_2+1} X_{\text{skip},\frac{n+1-l-j_2}{2}},
\end{equation}
note that the summation range of $j_2$ is dependent on $j_1$. To derive a finite summation bound for $j_2$, solve the inequality system:
$$
\begin{cases}
    0 < j_1 = \frac{n+1-l-j_2}{2} \\
    0 < j_2 \\
    j_2 < j_1 = \frac{n+1-l-j_2}{2} \\
\end{cases} \Leftrightarrow
\begin{cases}
    j_2 < n+1-l \\
    0 < j_2 \\
    j_2 < \frac{n+1-l}{3} \\
\end{cases}
$$
Apply these new summation ranges, and add conditional function $I$ ensuring all coefficient indices are integer. We now have the nth order coefficient for the expression: 
\begin{equation}
    G_n(\epsilon) = \sum_{j_2=0}^{(n+1-l)/3} I(j_2) j_2 X_{\text{diff},j_2+1} X_{\text{skip},\frac{n+1-l-j_2}{2}},
\end{equation}
where
\begin{equation}
I(j)=\begin{cases}
      1 & (\frac{n+1-l-j_2}{2}\text{ is an integer})\\
      0 & (\frac{n+1-l-j_2}{2}\text{ is not an integer}).\\
    \end{cases}
\end{equation} 
Repeating the procedure over all terms in an equation yields the general formula of its ordered equation set.

\subsection{Empirical identification of filter frequency}\label{appendix:low_pass}
\begin{figure}
    \centering
    \begin{subfigure}{0.45\textwidth}\label{fig:truncation1}
        \centering
        \includegraphics[width=\textwidth]{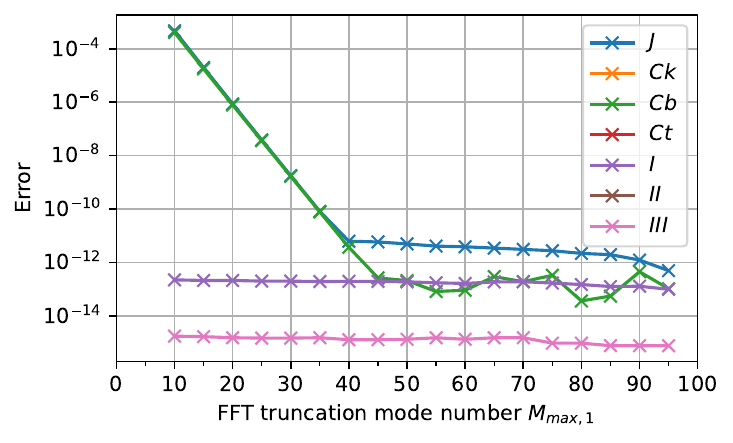}
        \caption{Cut-off at $M_{\text{max}, 1} = 45$}
    \end{subfigure}
    \begin{subfigure}{0.45\textwidth}
        \centering
        \includegraphics[width=\textwidth]{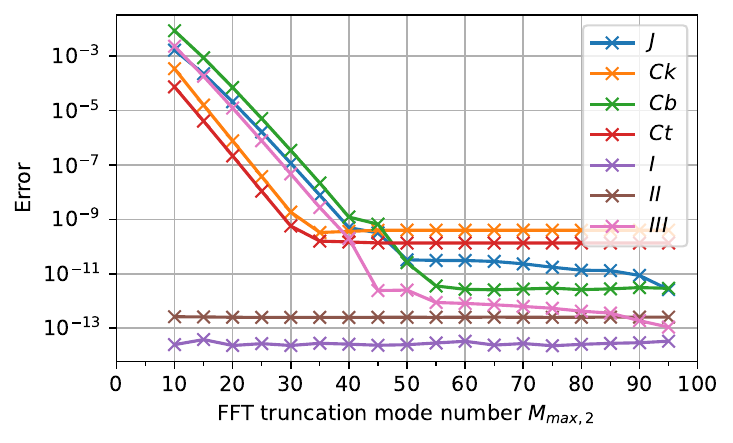}
        \caption{Cut-off at $M_{\text{max}, 2} = 50$}
    \end{subfigure}
    
    \begin{subfigure}{0.45\textwidth}
        \centering
        \includegraphics[width=\textwidth]{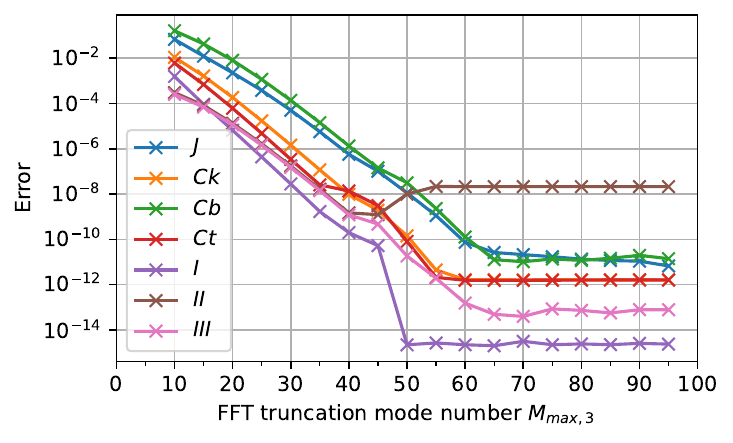}
        \caption{Cut-off at $M_{\text{max}, 3} = 45$}
    \end{subfigure}
    \begin{subfigure}{0.45\textwidth}
        \centering
        \includegraphics[width=\textwidth]{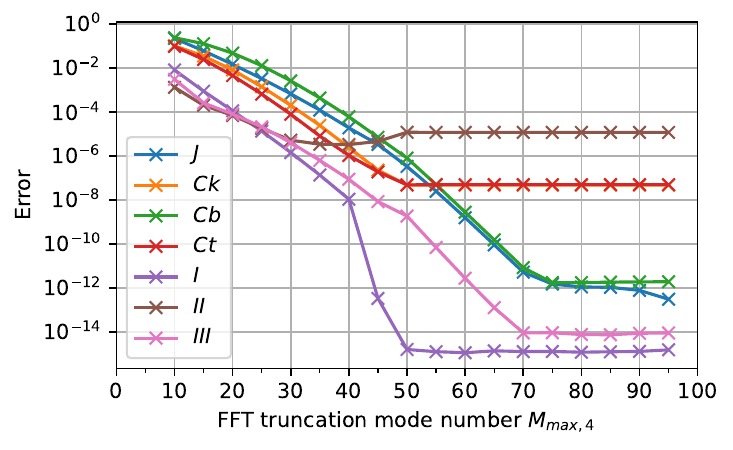}
        \caption{Cut-off at $M_{\text{max}, 4} = 40$}
    \end{subfigure}
    
    \begin{subfigure}{0.45\textwidth}
        \centering
        \includegraphics[width=\textwidth]{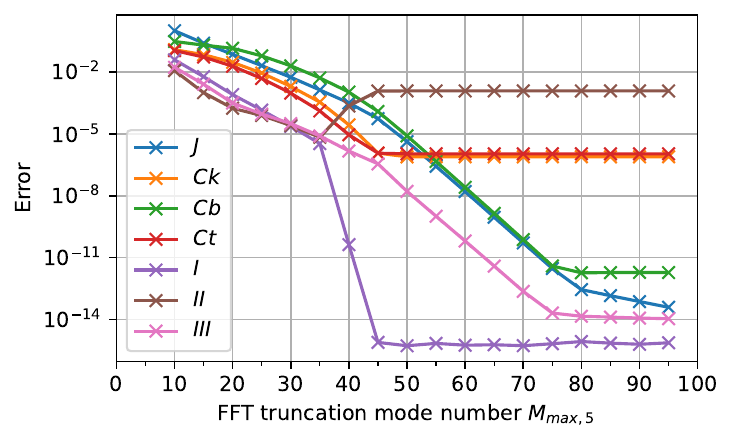}
        \caption{Cut-off at $M_{\text{max}, 5} = 35$}
    \end{subfigure}
    \begin{subfigure}{0.45\textwidth}\label{fig:truncation6}
        \centering
        \includegraphics[width=\textwidth]{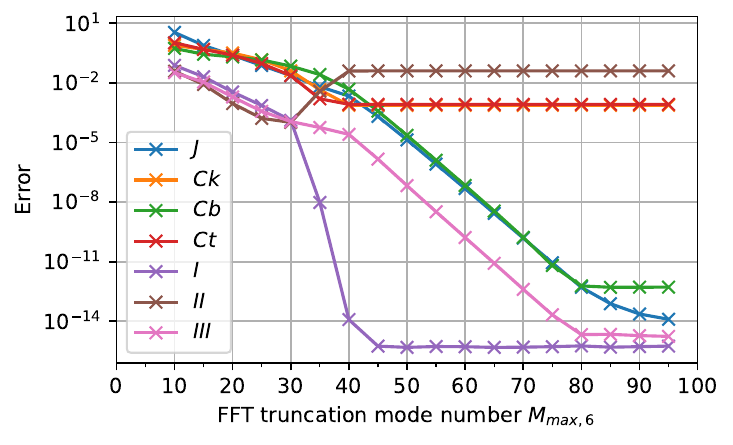}
        \caption{Cut-off at $M_{\text{max}, 6} = 30$}
    \end{subfigure}
    
    \caption{A 3-row by 2-column grid of plots.}\label{fig:truncations}
\end{figure}

Fig. \ref{fig:truncations} plots the iteration error curves at each order, as functions of the number of FFT modes preserved by the low-pass filter. The iterations are measured with the same method as in Fig. \ref{fig:exp}. We choose the truncation mode number $M_{\text{max}, n}$ at the "elbow" point of each plot, where the error becomes flat. Intuitively, keeping more modes than these "elbow points" may include more high-frequency noise without significantly reducing iteration error. 

\subsection{Method for calculating $\psi_{\text{crit},n}$}\label{appendix:psi_crit}
In \cite{landreman_figures_of_merit_2021}, $\psi_\text{crit}$ is solved by analytically finding the smallest $\psi$ that can satisfy:
\begin{equation}\label{eq:psi_crit_eq}
    \sqrt{g}\equiv\frac{\partial\textbf{r}}{\partial\psi}\cdot\left(\frac{\partial\textbf{r}}{\partial\chi}\times\frac{\partial\textbf{r}}{\partial\phi}\right) = 0
\end{equation}
using its closed-form formula in the $O(\epsilon^3)$ expansion by Landreman and Sengupta \cite{landreman_nae}. For higher orders, a similar analytical solution is hard to reach. Therefore, in \texttt{pyAQSC}, we estimate the value of $\psi_{\text{crit},n}$ numerically. 

$\psi_{\text{crit},n}$ can be interpreted as the smallest non-zero solution to the equation:
\begin{equation}
    \min_{\chi, \phi}\left[\sqrt{g_n}(\psi, \chi, \phi)\right]=0,
\end{equation}
where
\begin{equation}
    \sqrt{g_n}\equiv\frac{\partial\textbf{r}_n}{\partial\psi}\cdot\left(\frac{\partial\textbf{r}_n}{\partial\chi}\times\frac{\partial\textbf{r}_n}{\partial\phi}\right)
\end{equation}
This is tricky with Lagrangian multipliers. Because $\sqrt{g_n}(\psi=0)=0$ always holds, numerically solving
\begin{equation}
    \mathcal{L} = \psi + \lambda \min_{\chi,\phi}\sqrt{g_n}
\end{equation}
with Newton's method will often yield $\psi=0$.

In \texttt{pyAQSC}, we assume that the equilibrium is always sufficiently well-behaved, so that $\min_{\chi, \phi}\sqrt{g_n}\geq0$ for $\psi\in[0, \psi_{\text{crit},n}]$, and $\min_{\chi, \phi}\sqrt{g_n}<0$ for $\psi_{\text{crit}, n}$. Physically, this means that the volume enclosed by $\psi\leq\psi_{\text{crit},n}$ is non-zero, and that the flux surface can become well-behaved again for some $\psi\geq\psi_{\text{crit},n}$. With these two assumptions, the problem is simplified to the search for the unique zero intercept where $\min_{\chi,\phi}\sqrt{g_n}$ changes sign from positive to negative. 

In \texttt{pyAQSC}, we estimate $\min_{\chi,\phi}\sqrt{g_n}$ by evaluating $\sqrt{g_n}$ on a uniform grid, and search for this intercept by bisection. We choose the left starting point at $\psi=0$, and the right starting point at 
\begin{equation}
    \psi_\text{init}\equiv \frac{l}{2\pi}\sqrt{B_0}.
\end{equation}
This is a crude estimate of $\psi$ where the distance from the axis equals to the effective major radius, $l/2\pi$. In the case where $\left[\min_{\chi,\phi}\sqrt{g_n}\right](\psi_\text{init})>0$,  $\psi_\text{init}$ is multiplied by 2 until $\left[\min_{\chi,\phi}\sqrt{g_n}\right](\psi_\text{init})<0$.
\end{appendices}

\printbibliography

\end{document}